\documentclass[mathleft
]{an}
\usepackage{graphicx,booktabs}
\usepackage{times}
\begin{document}

\Pagespan{1}{}
\Yearpublication{2007}%
\Yearsubmission{2006}%
\Month{}%
\Volume{???}%
\Issue{???}%

\title{The inner structure of the S0 galaxy NGC\,3384}

\author{H. Meusinger\inst{1}\fnmsep\thanks{Corresponding author:
  \email{meus@tls-tautenburg.de}\newline},
        H. A. Ismail\inst{2},
	\and P. Notni\inst{3}
\thanks{Visiting astronomer,  German-Spanish Astronomical Center, Calar Alto, 
operated by the MPIA, Heidelberg, jointly with the Spanish National Commission 
for  Astronomy.}
}
\titlerunning{NGC3384}
\authorrunning{H. Meusinger, H. Ismail, P. Notni}
\institute{
Th\"uringer Landessternwarte, Sternwarte 5, D-07778 Tautenburg, Germany
\and National Research Institute of Astronomy and Geophysics, Cairo, Egypt
\and Leo-Sachse-Str.97, 07749 Jena, Germany}

\received{28 Dec 2006}
\accepted{30 Jan 2007}
\publonline{later}

\keywords{galaxies: peculiar -- galaxies: individual (NGC3384) --
galaxies: structure -- galaxies: evolution}

\abstract{%
We re-investigate the lenticular galaxy NGC 3384, a member of the Leo~I
galaxy group, using {\it HST} and multi-colour Calar Alto observations. 
Various approaches are used to visualize the two  
known peculiar components, the so-called inner component (IC) and the
elongated component (EC), on the {\it HST} images. The methods were checked 
in detail using synthetical images from simulated 
galaxies. For the first time, we make the IC as well as the inner part of 
the EC visible on direct images. The results confirm both 
the bar-like appearance of the inner EC and the quasi-elliptical shape of the 
IC. The IC resembles an inclined disk where the surface brightness becomes 
successively shallower towards the centre compared to
an exponential profile. The orientation of the inner part of the EC 
is perpendicular to the major axis of the IC. The broad-band colour 
indices are shown to 
be in agreement with model predictions for a 5 to 7\,Gyr old 
stellar population of quasi-solar metallicity. No significant large-scale
variations of the colour indices over the main body of the galaxy are found. 
We discuss the previously reported colour gradients close to the nucleus 
and argue that the most plausible explanation is reddening by 
small amounts of dust though unsharpe masked {\it HST} images do not 
provide significant hints for clumpy dust. According to the 
episodic dust settling scenario suggested by Lauer et al (\cite{Lauer05}),
the very low dust fraction indicates that NGC\,3384 
is in a post-activity phase and at the beginning of a new dust-settling cycle.
}

\maketitle

%
\section{Introduction}\label{intro}
%

NGC\,3384 is an apparently unspectacular yet fascinating  ``classical'' 
S0 galaxy in the nearby galaxy group Leo\,I. 
Its appearance on a photographic plate 
was described by Pease (\cite{Pease20}) as 
``an astonishing nebula consisting of a bright center $40\arcsec$ 
diameter, on which are superimposed a very bright elongated nucleus 
$19\arcsec \times 10\arcsec, p\ 45\degr$, crossed by a second bright 
nucleus $40\arcsec \times 5\arcsec, p\ 130\degr$, presenting a 
Saturn-like appearance". Since that time, NGC\,3384 was 
a target of a variety of photometric, spectroscopic, and kinematic 
studies of samples of 
nearby early-type ga\-laxies (e.g.,  Fisher \cite{Fisher97}; Pinkney et al. 
\cite{Pinkney03}; Emsellem  et al. \cite{Emsellem04}; Erwin \cite{Erwin04}; 
Lauer et al. \cite{Lauer05}; Sil'\-chenko \cite{Silchenko06}). 
Early-type galaxies are now known to exhibit a rich variety of 
distinct morphological features like e.g., twisted and/or non-concentric 
isophotes, disky or boxy deviations of the isophotes from pure ellipses 
(Lauer \cite{Lauer85}), outer shells (Malin \& Carter \cite{Malin83}), 
pseudo-bulges (Kormendy \& Kennicutt \cite{Kormendy04}),
double bars (Erwin \cite{Erwin04}), and supermassive central black holes 
(Kormendy \& Richstone \cite{Kormendy95}).

All these peculiar features have been identified in NGC 3384:
strong isophote twist and centre position variation (Barbon et al. 
\cite{Barbon76}), outer boxy and inner disky isophotes (Busarello 
et al. \cite{Busarello96}); a faint extended luminous arc (outer shell)
at the N-E side and outside the main galaxy was first discovered 
by Malin (\cite{Malin84}) from a photographically amplified image and 
is clearly indicated also on the blue POSS\,II image. NGC\,3384 was 
classified by Pinkney et al (\cite{Pinkney03}) as a possible pseudo-bulge
galaxy and by Erwin (\cite{Erwin04}) as a bar-in-bar galaxy. 
The dynamically derived mass of the central black hole is 
$1.6\,10^7 \mathrm{m}_{\odot}$ (Tremaine et al \cite{Tremaine02}).
Finally, we note that NGC\,3384 is one of the only two galaxies were
a new class of extended star clusters known as ``faint fuzzies'' were 
detected so far (Burkert et al \cite{Burkert05}). 

The interest in such morphological peculiarities is main\-ly motivated by 
the fact that they are tracers of the past galaxy evolution, either rapid
evolution driven by mergers (Toomre \& Toomre \cite{Toomre72}) or slow, 
secular evolution driven by other 
external or internal processes (Kormendy \& Kennicutt \cite{Kormendy04}). 
NGC\,3384 is a good example for the investigation of such processes 
in the relatively poor galaxy environment of a loose galaxy group. 
A past collision of NGC\,3384 with NGC\,3368 was 
suggested by Rood \& Williams (\cite{Rood85}) whereas the detailed study of 
the Leo\,I group galaxies by Sil'chenko et al (\cite{Silchenko03}) suggests
that many of their properties are in line with the scenario of tidal 
interactions of the galaxies with a supergiant intergalactic ring of HI clouds.  

A detailed study of NGC\,3384 was presented by Busa\-rello et al 
(\cite{Busarello96})
where also a comprehensive summary of its basic properties can be found. 
Their investigation was based on imaging and spectroscopy from 
ground-based telescopes, including the ESO NTT. The NTT image, processed 
with an adaptive filter, revealed the existence of an hitherto
unknown, small ($r \sim 6\arcsec$) central component aligned to the
major axis of the galaxy ($PA \approx 46\degr$). 
This is called by of Busarello 
et al the {\it inner component} (IC). Kinematical data from long-slit 
or integral-field spectroscopy are in line with the interpretation 
of the IC as a very rapidly rotating component which is kinematically colder 
than the surrounding (Busarello et al \cite{Busarello96}; 
Fisher \cite{Fisher97}; Sil'chenko et al \cite{Silchenko03}; Emsellem et 
al \cite{Emsellem04}). The previously known almost peanut-shaped 
{\it elongated component} (EC), at position angle $PA \approx 132\degr$, 
is probably best interpreted as a bar. 
The EC has been known for a long time (Pease \cite{Pease20}). 
For the IC, on the other hand, no direct image was extracted so far 
due to its small size and intrinsic faintness. 

Busarello et al mentioned 
that their re-investigation of NGC\,3384 was relying on the rule 
``that it is far easier to see something when it is known to be there''. 
Following the same rule, the present paper is aimed mainly at the 
visualization of the IC and the discussion of its properties
based on new, high-resolution images. A second aim is to characterize the 
stellar population in the galaxy by the comparison of colour indices from
multi-colour photometry with population synthesis models.

The paper is organized as follows. First, we give a brief overview over the 
observational data in Sect.\,2, than we pre\-sent large-scale light
profiles in Sect.\,3. In Sect.\,4, we analyze the structures in the inner 
$\sim 15\arcsec$ on {\it HST} images. Sections 5, 6, and 7 deal with 
the comparison of the colour data with evolution models and the 
possible presence of dust and, finally, the conclusions. Throughout this 
paper we assume a distance of $d = 11.7$\,Mpc (Tonry et al \cite{Tonry01}), 
corresponding to a scale of 56.7\,pc per arcsec.\\

%
\section{Observational data}\label{observation}
%

Whereas the study of the small-scale structure in the centre of the galaxy
clearly requires high spatial resolution, the sky background correction is 
essential for the investigation of the colour distribution.
Lauer et al (\cite{Lauer05}) have measured a pseudo-sky level from the 
corner of the WFPC2 WF3 CCD at a distance of about 1\farcm8 from the 
centre i.e., still within the galaxy.  A major aim 
of the present study is to derive ages of the galaxy components 
from the projected distribution of colours. The results are very sensitive 
to the photometric calibration and in particular to the correct
sky background subtraction. Therefore, the field of view must be larger 
than the angular size of the galaxy which is about $5\farcm5 \times 2\farcm5$.

For the study of the colour distribution in NGC\,3384 we prefer ground-based 
observations with a sufficiently large field of view, despite a 
poorer spatial resolution.
CCD observations were obtained by one of us (P.N.) in February 1995 using 
the 1.23\,m telescope on Calar Alto, equipped with a TEK CCD camera with
$1024 \times 1024$ pixels of $24\mu\mbox{m} \times 24\mu\mbox{m}$. The 
field of view amounts to 8\farcm5 $\times$ 8\farcm5 with an image
scale of 0\farcs502 pixel$^{-1}$. Images were taken 
 through Johnson-Cousins B, V, R and I filters with 
typically two or more exposures per filter. Typical exposure times 
are 1\,500\,s U, 500\,s B, 300\,s V, 100\,s R, and 100\,s I.
The sky conditions were good, the seeing varied between 1\arcsec and 
2\arcsec. Flatfields were obtained during morning and evening twilight, 
from which a master flat has been made. 

The bias-corrected and flatfielded frames were matched carefully in 
position using the foreground stars in the field, and transformed to a 
common coordinate system. Cosmics were eliminated as usual by  
interpolation between neighbouring pixels. A similar procedure
was used to clean the image from bright foreground stars 
for precise surface photometry. Finally, all images from 
the same filter band were co-added to produce one final image per band.
All image reductions have been conducted using standard procedures 
from the ESO MIDAS image processing package. 

High resolution images of the inner part of NGC\,3384 are available from the 
data archives. {\it Hubble Space Telescope (HST)} images were taken in 1995 
with the planetary camera CCD in WFPC2  (PI: T. Lauer; see Lauer et al 
\cite{Lauer05}). 
The field of view amounts to $36\arcsec \times 36\arcsec$ with an image scale 
of 0\farcs0456 px$^{-1}$, corresponding to 2.58\,pc px$^{-1}$ at the distance 
of NGC 3384. Three exposures were taken in each of the filter bands 
F555W and F814W, equi\-valent to broadband V and I, respectively
with 350\,s per exposure. Hence, cosmic-ray event repair could be done easily by the 
comparison of the frames from the same filter band.  In addition, we use 
on-the-fly-reprocessed ACS/HRC images, taken in June 2005 (PI: R. O'Connell), 
downloaded from the Multimission Archive at STScI (MAST). The exposure times are 
2552\,s, 1912\,s, and 432\,s, respectively, for the three filters F250W, F330W, 
and F555W. At higher energies ($0.3 \ldots 8$ keV), a 10\,ks {\it CHANDRA} event 
file, taken in Oct. 2004 (PI: Y. Terashima) with the S-array 
of the Advanced CCD Imaging Spectrometer (ACIS), is available from the 
{\it CHANDRA} Data Archive; the pixel scale is 0\farcs492\,px$^{-1}$.

%
\section{Global properties from the BVR surface photometry}\label{UBVR}
%

In this Section, we use the UBVR frames from Calar Alto because 
they cover the whole galaxy. Each frame was thoroughly background-corrected 
before a final image per filter band was produced.
The brightness of the sky background 
in each frame was estimated using polynomials of  second order for 
the ``average'' value in selected background areas. 
The areas are chosen via small boxes distributed over the 
outskirts of the frames. Then an average image for each filter is obtained.
 A convenient compilation of aperture photoelectric magnitudes of galaxies 
is published by Prugniel \& Heraudeau (\cite{Prugniel98}), which have been used here 
to transform our instrumental magnitudes to the standard Johnson-Cousins system. The 
rms error values of our photometric zero points in B, V, R and I are 0.03, 
0.01, 0.02 and 0.01 mag, respectively. 

The quantitative analysis of the surface photometry was obtained by 
fitting the isophotes of the galaxy with concentric ellipses using the task 
ELLIPSE of the ISOPHOTE package in IRAF.  The task reads the 
two-dimensional image and produces a table containing the essential parameters for each
fitted isophote such as the brightness level, ellipticity, position angle, etc. 
The table content is used later on as input in the IRAF task BMODEL to 
reconstruct a galaxy image with perfect elliptical and concentric isophotes.

\begin{figure}[hhhh]
\includegraphics[width=5.5cm,angle=270,clip=true]{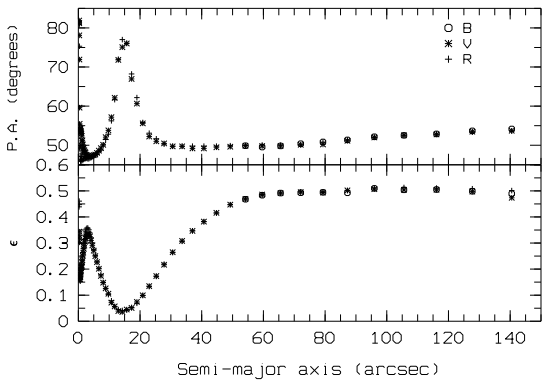}
\caption{The change of the position angle (top) and the ellipticity 
(bottom) with the semi-major axis in the B, V, and R band.}
\label{pos_ell}
\end{figure}

\begin{figure}[hhhh]
\includegraphics[width=5.5cm,angle=270,clip=true]{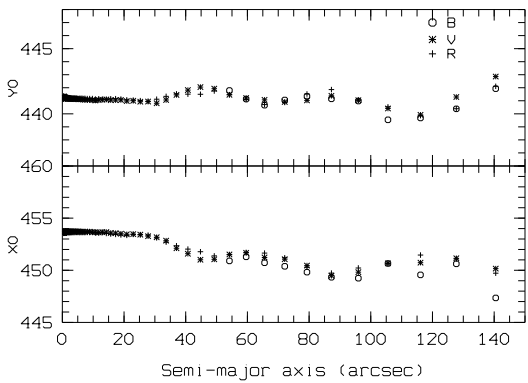}
\caption{The drift of the centre coordinates (pixels).  
}
\label{centre}
\end{figure}

\begin{figure}[hhhh]
\includegraphics[width=3.8cm,angle=270,clip=true]{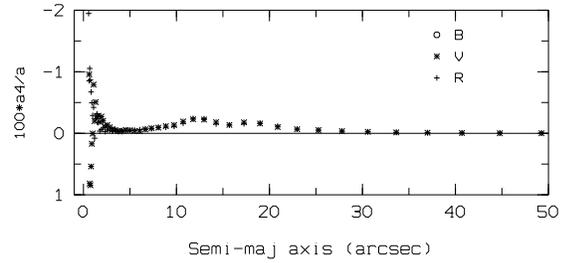}
\caption{The change of the normalized Fourier parameter $a_4/a$.  
}
\label{a4}
\end{figure}

The results based on the ellipse fitting are shown in Fig.s \ref{pos_ell} to \ref{a4}.
Figure \ref{pos_ell} gives the variation of the position angle and the ellipticity  
$\epsilon = 1 - b/a$, where $a$ and $b$ are the semi-major axis and the semi-minor axis, 
respectively, as a function of $a$. The ellipticity has a local maximum at 
$a \approx 3\arcsec$  where the position angle reaches a minimum, and vice versa the 
maximum of the position angle at $a \approx 15\arcsec$  corresponds to a minimum of the 
ellipticity. The same features have been reported already by Busarello et al
(\cite{Busarello96}) in the B and R bands and by Sil'chenko et al 
(\cite{Silchenko03}) in the JHK bands. The estimated mean values ($\pm$ standard 
deviation) of the ellipticity and the position angle at $a = 50\arcsec$ are 
$\epsilon_{50} = 0.49$ and $PA_{\, 50} = 50.6^\circ$, respectively, 
in perfect agreement with Busarello et al and Sil'chenko et al. 
These authors interpreted the local maximum of the ellipticity  in the inner 
region ($a \approx 5\arcsec$) as due to the presence of the IC. The minimum of 
the ellipticity 
and the corresponding maximum of the position angle at $a \approx 15\arcsec$ 
is assigned to the superposition of a flattened spheroid, $PA \approx 50\degr$, 
with the EC which is elongated almost along the minor axis. 
       
The variation of the centres of the isophotes is shown in Fig.\,\ref{centre}
where $x0, y0$ increase toward W and N, respectively.  
Again, the general trend is similar to the results shown by Busarello et al.
Especially, we establish their finding that the drift of the isophote centre 
is very small within the region of the EC  ($\la 25\arcsec$ along the minor 
axis). At larger distances, the isophote centres shift toward NE, opposite 
to the position of the neighbour galaxy NGC\,3379, indicating the large-scale 
asymmetry discussed by Busarello et al. The finding that 
there is also a NE shift
at $\sim 110\ldots 140\arcsec$ is original to the present paper.

Low-order deviations of the isophotes from perfect ellipses are 
quantified by the fourth order Fourier harmonics $a_4$ with $a_4>0$ for disky 
and $a_4<0$ for boxy isophotes. The boxy shape of NGC\,3384 is clearly 
seen at $a \approx 10\arcsec \ldots$  $20\arcsec$\, in Fig.\,\ref{a4}. 
A disky inner structure is marginally indicated in the B image at 
$a \approx 3\arcsec$.

%
\section{Inner substructure}\label{IS}
%

%
\subsection{Surface photometry}
%

For the analysis of the inner substructure we started with the PC part of the 
{\it HST} WFPC2 images. Following Lauer et al ({\cite{Lauer05}), we first deconvolved the 
PSF using the Lucy-Rich\-ardson algorithm \ (MIDAS \ procedure \ DECONVOL\-VE/  FLUCY). 
Except for some reddening ($\Delta (V-I) \approx 0.1$\,mag towards the centre 
within the innermost 10\arcsec, see also Lauer et al (\cite{Lauer05}, 
their Fig.\,3),
the images in the F555W filter and the F814W filter are very similar.  
Therefore, we decided to co-add both frames to get a deeper combined image with 
increased signal-to-noise ratio in the outer parts. In addition, the
{\it HST} ACS images were analyzed in a similar way but without co-adding the 
frames from the three different filter bands.

\begin{figure}[hhhh]
\includegraphics[width=6.0cm,angle=0,clip=true]{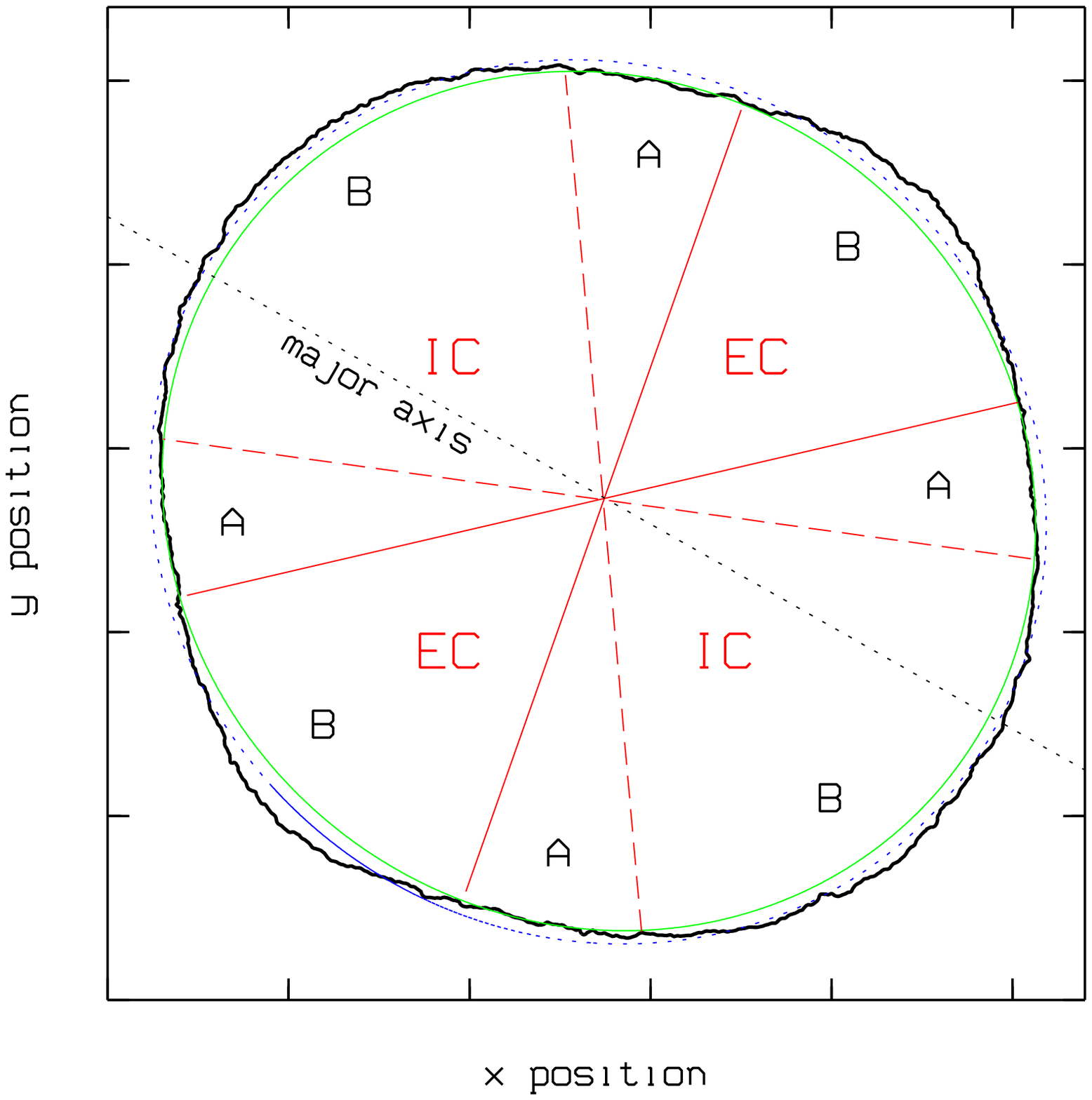}
\caption{
A single outer isophote from the PC part of the {\it HST} WFPC2 image (thick
curve) clearly showing deviations from ellipticity due to the 
``inner component'' (IC) and the ``elongated component'' (EC).
Best fitting ellipses are shown as dotted curve (EF = ellipse fitting to the 
whole isophote) and solid curve (IEF = {\it inner} ellipse fitting, i.e. 
in the angular intervals between IC and EC only), respectively. 
The direction of the major axis  of the fitting ellipse ($a = 11\farcs2$, 
PA$=60\degr$) is indicated by the dotted line. For the meaning of the labels 
``A'' and ``B'' see Sect.\ref{methods}.}
\label{isophote}
\end{figure}

\begin{figure}[hbpt]
\includegraphics[width=7.8cm,angle=0,clip=true]{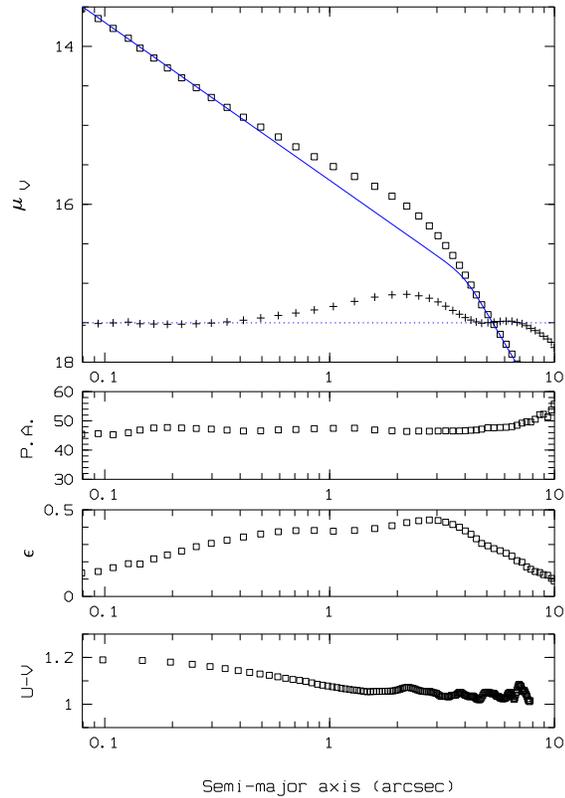}
\caption{Surface photometry for the {\it HST} ACS F555W filter. The style is similar
to that in Lauer et al (2005). The solid line in the top panel is the Nuker law fitted to 
the brightness profile; the trace of the bottom of the panel (crosses) shows the 
residuals from the fit (the dotted line indicates the zero level). The next 
panels give the isophote position angle, the ellipticity $\epsilon = (1-b/a)$, and 
the colour index $U - V$  as functions of semi-major axis. The colour index is measured 
in a $1\arcsec$ wide strip along the major axis. 
As in Lauer et al, no Galactic absorption correction has been applied.}
\label{Nuker}
\end{figure}

The isophote plot in Fig.\ref{WFPC2} indicates a boxy shape in the 
outer part and a disky shape in the inner part in agreement with the 
Fourier coefficient $a_4$ from the Calar Alto images
(Fig.\,\ref{a4}; see also Busarello et al \cite{Busarello96}). 
According to Lauer (\cite{Lauer85}), boxies are described by ``pulling the 
iso\-phote in towards the centre of the galaxy on the major and minor axes, 
but pushing it out 45\degr\, in between''. The closer inspection of the 
isophotes (see Fig.\,\ref{isophote}) reveals that their shape is slightly 
different from pure boxies in the sense of Lauer. It is reasonable to interpret 
the isophote shape seen in Fig.\,\ref{isophote} as due to the superposition of 
the IC and the EC. A disky inner structure, caused by the IC, is indicated by 
weak ansae at $PA \approx 47\degr$. Though not conspicuous in the 
contour plot in Fig.\,\ref{WFPC2}, they are clearly indicated by the
results of the ellipse fit at $a \approx 3\ldots6\arcsec$. 
For the EC, with a diameter of about 20\arcsec$\ldots$40\arcsec, only the 
inner part is covered by the field of the planetary camera CCD. 

In Fig.\,\ref{Nuker}, we plot the surface photometry for the inner 
$10\arcsec$ on the {\it HST} ACS F555W image. As the style of the presentation 
is similar to the surface photometry given by Lauer et al (\cite{Lauer05}) 
for the {\it HST} WFPC2 F555W image, the perfect agreement with their 
results is easily recognizable. The IC is indicated by the noticeable 
increase of the ellipticity between $a \approx 0.2\arcsec$ and $5\arcsec$
with a maximum $\epsilon_{\mathrm max} = 0.44$, i.e.
$\sim 10$\% higher than in the NTT image from Busarello et al.
At small distances from the centre ($a \la 0\farcs1$), the nucleus is 
dominating. From the shape of the isophotes (Fig.\,\ref{WFPC2}), we conclude 
that the EC becomes important only at large distances ($r \ga 10\arcsec$ 
along the minor axis). 

It is well known that the inner surface brightness profiles of early-type 
galaxies can be approximated by a Nuker law (Lauer et al. \cite{Lauer95}), 
which consists essentially of two different power laws. In particular,
the Nuker fit is useful to distinguish between power-law galaxies and core 
galaxies (Lauer et al. \cite{Lauer05}). 
NGC 3384 is clearly characterized as a power-law galaxy by the Nuker fit, 
though the fit shown by Lauer et al is obviously not perfect (as, however, for
other galaxies too). We find it tempting to speculate that the deviations 
might be caused by the IC. In 
Fig.\,\ref{Nuker} we show a Nuker fit with parameters different
from those given by Lauer et al. (\cite{Lauer05}) but providing 
a good fit for the distance range where the IC is not dominating. 
The residuals from this fit show indeed a nice correlation with the 
ellipticity, as expected if the residuals are dominated by the IC. 


\begin{figure*}[h]
\begin{tabbing}
\fbox{\resizebox{4.3cm}{4.3cm}{\includegraphics[clip=true]{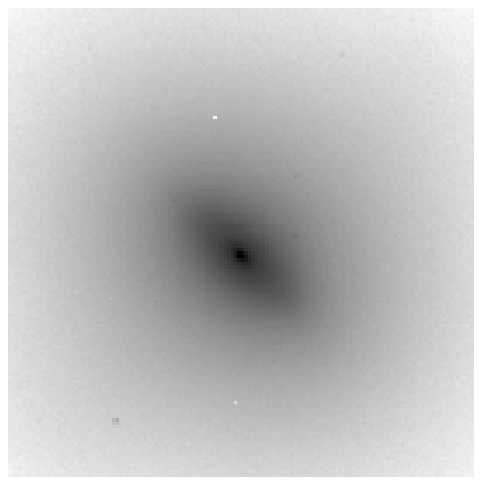}}} \hfill \=
\fbox{\resizebox{4.3cm}{4.3cm}{\includegraphics[clip=true]{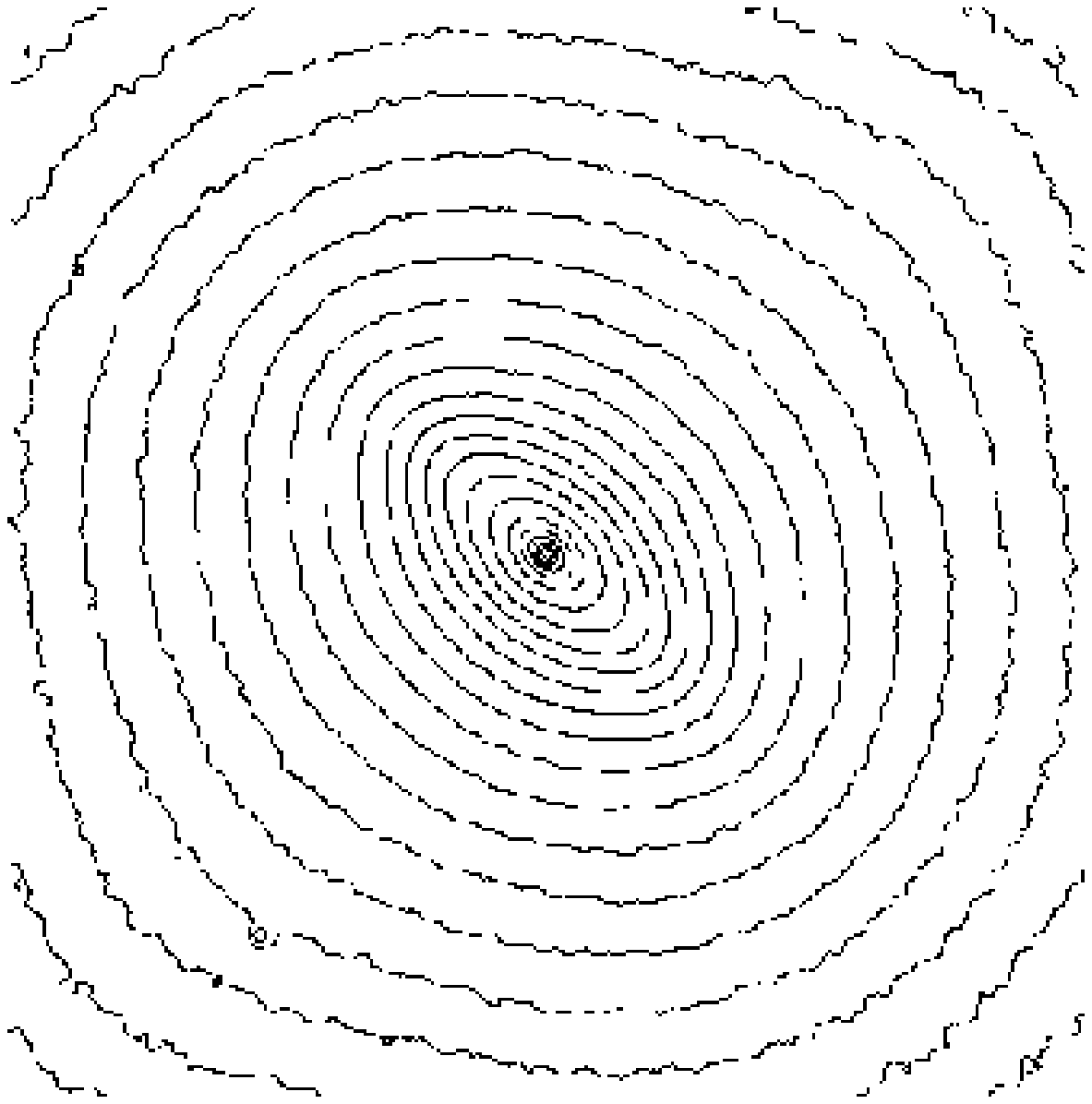}}} \hfill \=
\fbox{\resizebox{4.3cm}{4.3cm}{\includegraphics[clip=true]{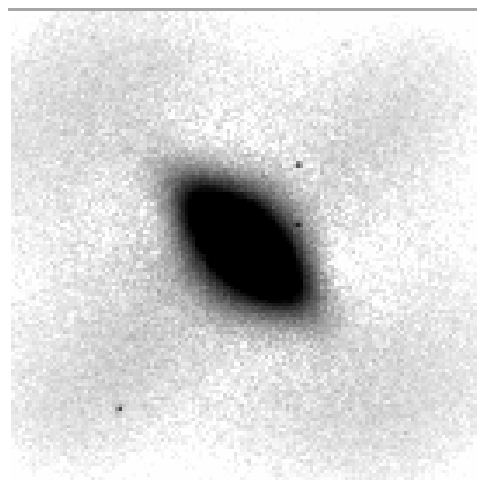}}} \= \\
\fbox{\resizebox{4.3cm}{4.3cm}{\includegraphics[clip=true]{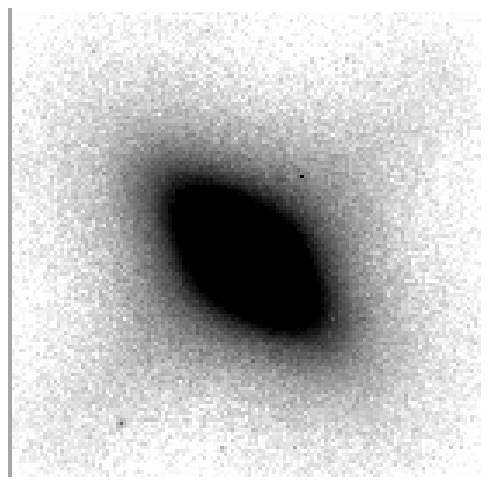}}}\>
\fbox{\resizebox{4.3cm}{4.3cm}{\includegraphics[clip=true]{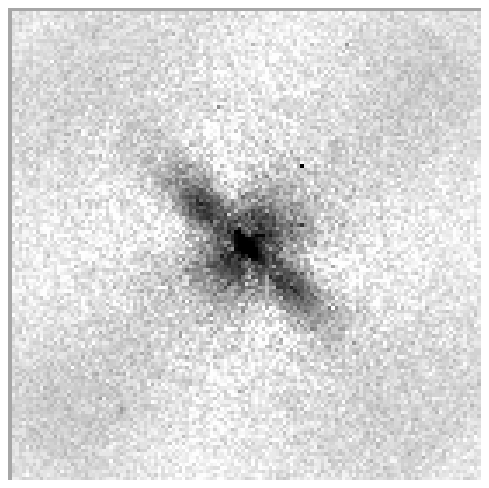}}}\> 
\fbox{\resizebox{4.3cm}{4.3cm}{\includegraphics[clip=true]{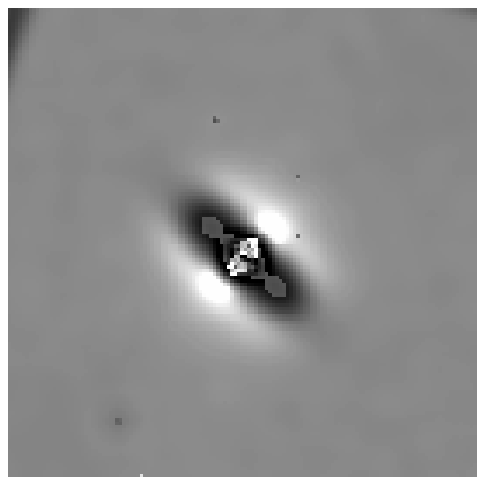}}}\> \\
\end{tabbing}
\caption{
Inverse logarithmic grey-scale presentation of the combined
{\it HST} planetary camera image of NGC\,3384.  
{\it Top, left to right}:
(a) original image,
(b) contour plot,
(c) residual image after subtraction of the profile fitting model.
{\it Bottom, left to right}: residual images after subtraction of
the corresponding model image derived from
(d) unsharp masking,
(e) inner ellipse fitting (IEF).   
(f) result from adaptive Laplace-filtering. 
Image size $24\arcsec \times 24\arcsec$ 
(corresponding to 1.36\,kpc at the distance of NGC\,3384). N up, E left.
}
\label{WFPC2}
\end{figure*}

\begin{figure*}[hhhh]
\begin{tabbing}
\fbox{\resizebox{4.0cm}{4.0cm}{\includegraphics[clip=true]{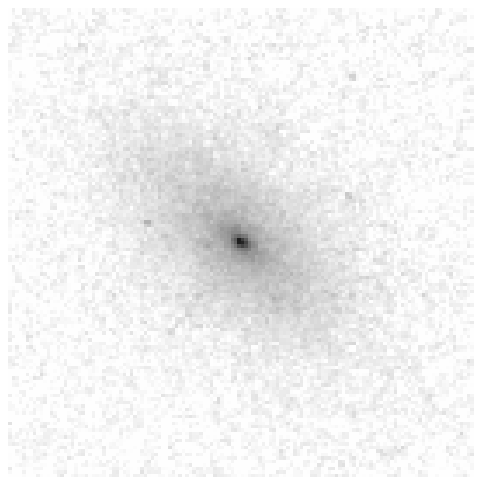}}} \hfill \=
\fbox{\resizebox{4.0cm}{4.0cm}{\includegraphics[clip=true]{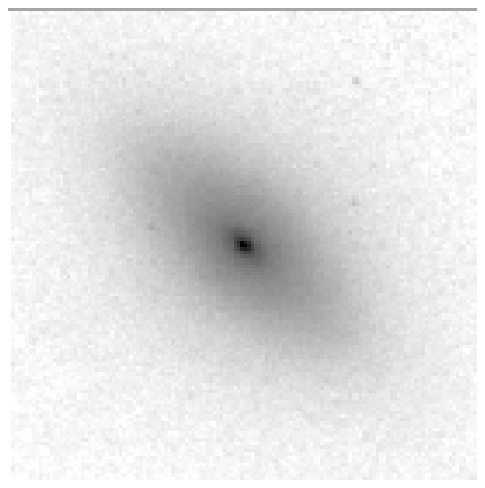}}} \hfill \=
\fbox{\resizebox{4.0cm}{4.0cm}{\includegraphics[clip=true]{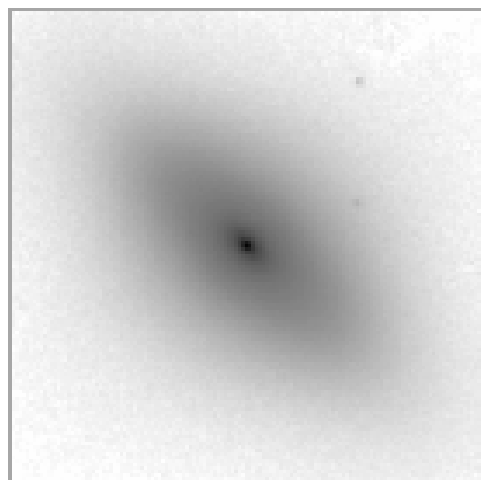}}} \= \\
\fbox{\resizebox{4.0cm}{4.0cm}{\includegraphics[clip=true]{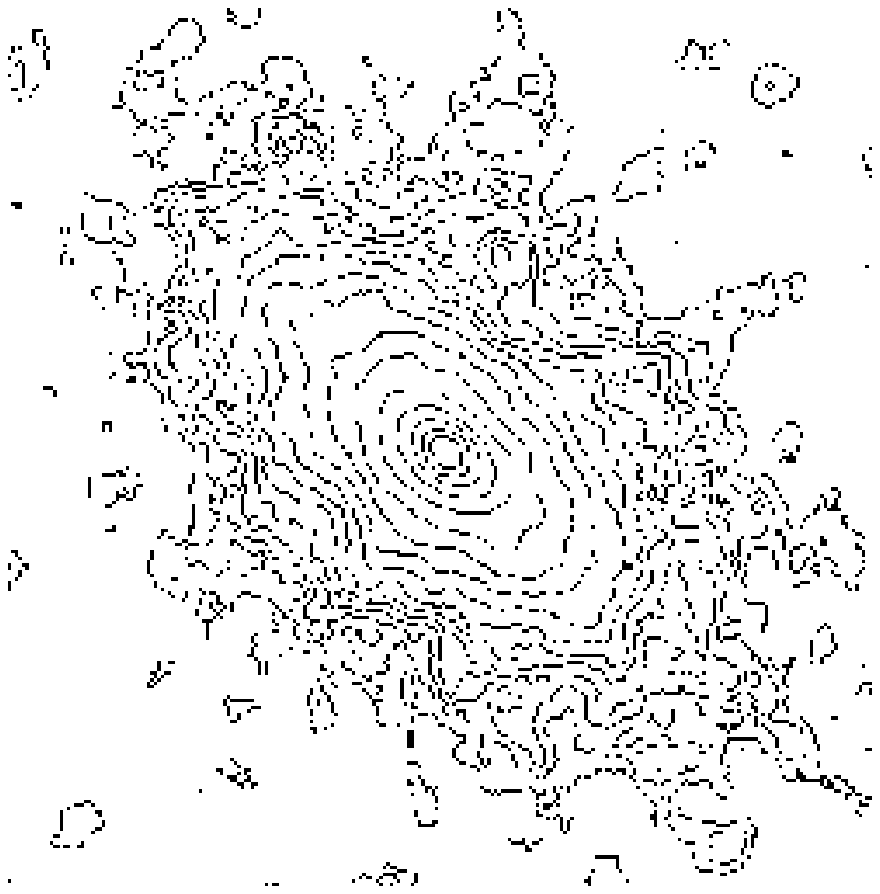}}} \>
\fbox{\resizebox{4.0cm}{4.0cm}{\includegraphics[clip=true]{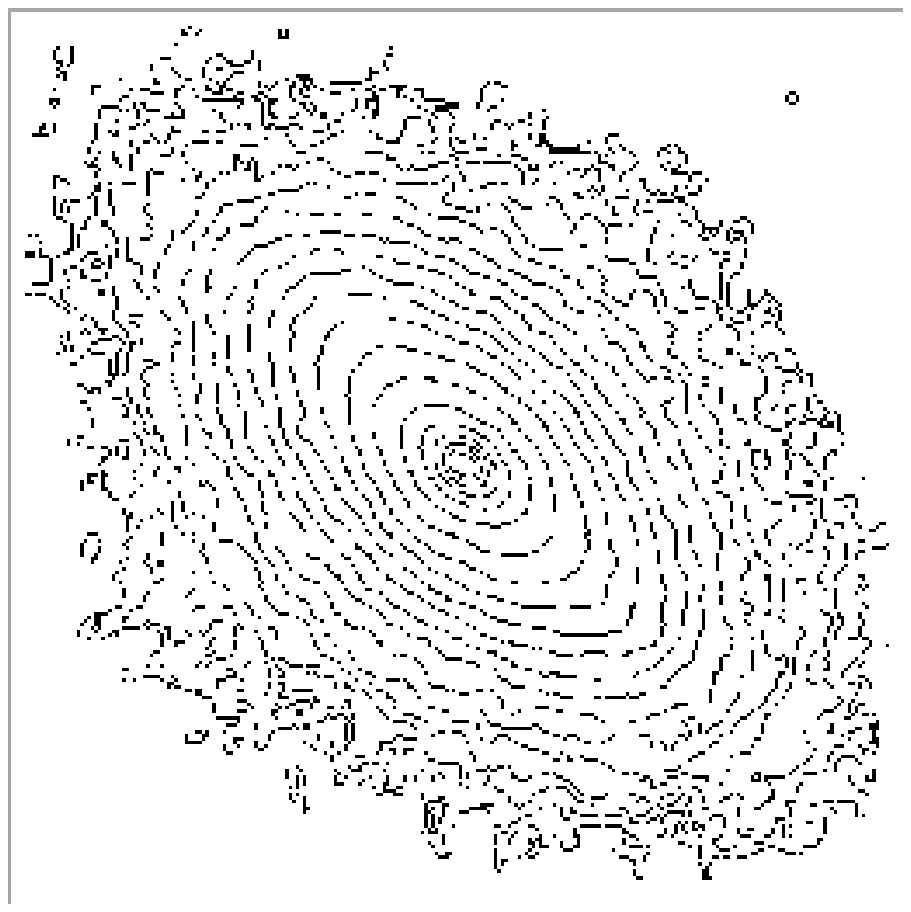}}} \>
\fbox{\resizebox{4.0cm}{4.0cm}{\includegraphics[clip=true]{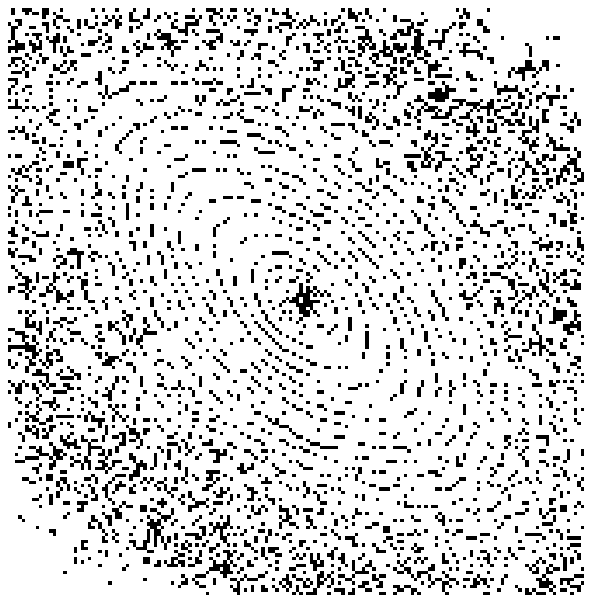}}} \> \\
\end{tabbing}
\caption{{\it Top:} 
the inner $12\arcsec \times 12\arcsec$ region of NGC\,3384 after subtraction of the median-filtered 
{\it HST} ACS images in the F250W, F330W, and F555W filter band (left to right). 
The intensity interval is the same, N up, E left. 
{\it Bottom:} corresponding contour plots (see text).
}
\label{ACS}
\end{figure*}

\newpage
\clearpage

\begin{figure*}[hbbb]
\begin{tabbing}
\fbox{\resizebox{4.0cm}{4.0cm}{\includegraphics[clip=true]{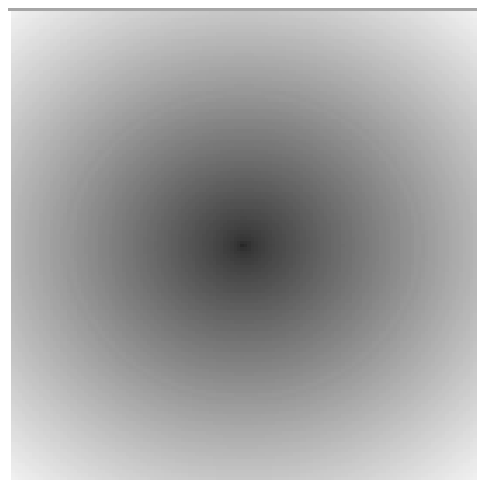}}} \hfill \=
\fbox{\resizebox{4.0cm}{4.0cm}{\includegraphics[clip=true]{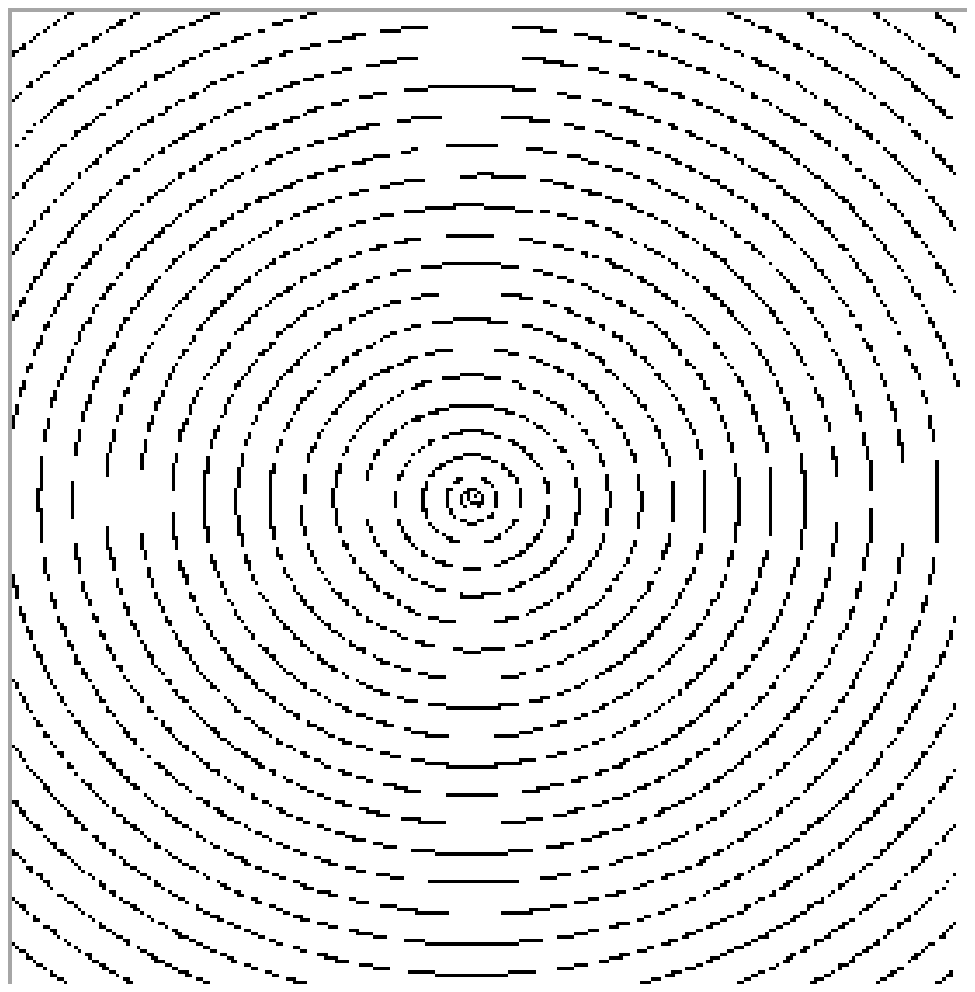}}} \hfill \= 
\fbox{\resizebox{4.0cm}{4.0cm}{\includegraphics[clip=true]{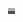}}} \= \\
\fbox{\resizebox{4.0cm}{4.0cm}{\includegraphics[clip=true]{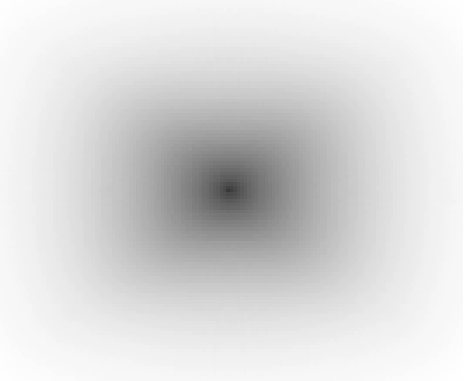}}}\>
\fbox{\resizebox{4.0cm}{4.0cm}{\includegraphics[clip=true]{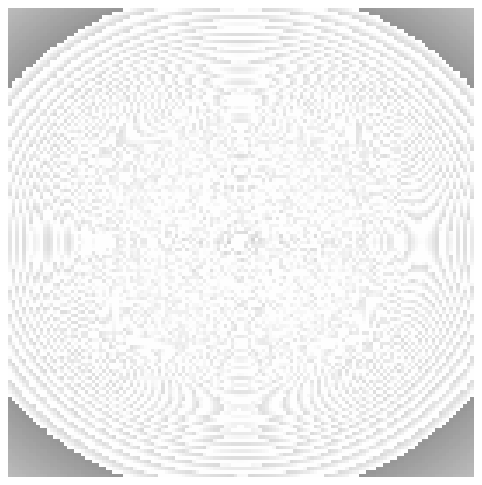}}}\> 
\fbox{\resizebox{4.0cm}{4.0cm}{\includegraphics[clip=true]{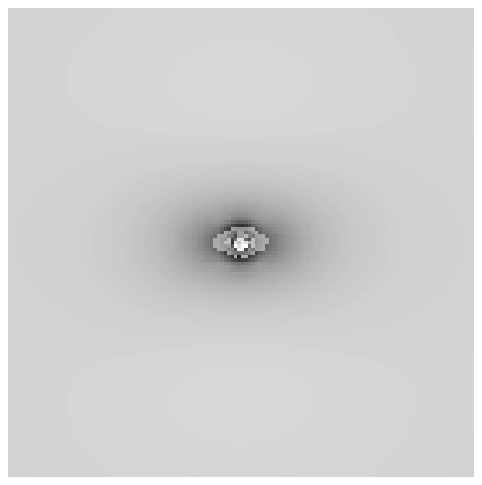}}}\> \\
\end{tabbing}
\caption{Inverse logarithmic grey-scale presentation of model {\bf DBN}. 
{\it Top, left to right:} 
(a) all components, 
(b) contour plot of image (a),
(c) components without disk and bulge (upper right).
{\it Bottom, left to right:}
(d) residual image after subtraction of the median-filtered image, and 
(e) residual image after subtraction of the corrected IEF model. 
(f) adaptive Laplace-filtered image.
Same scale for all images. 
}
\label{DBN}
\end{figure*}

\begin{figure*}[hbpt]
\begin{tabbing}
\fbox{\resizebox{4.0cm}{4.0cm}{\includegraphics[clip=true]{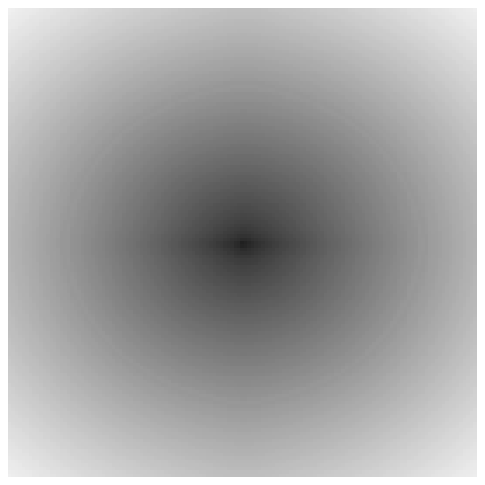}}} \hfill \=
\fbox{\resizebox{4.0cm}{4.0cm}{\includegraphics[clip=true]{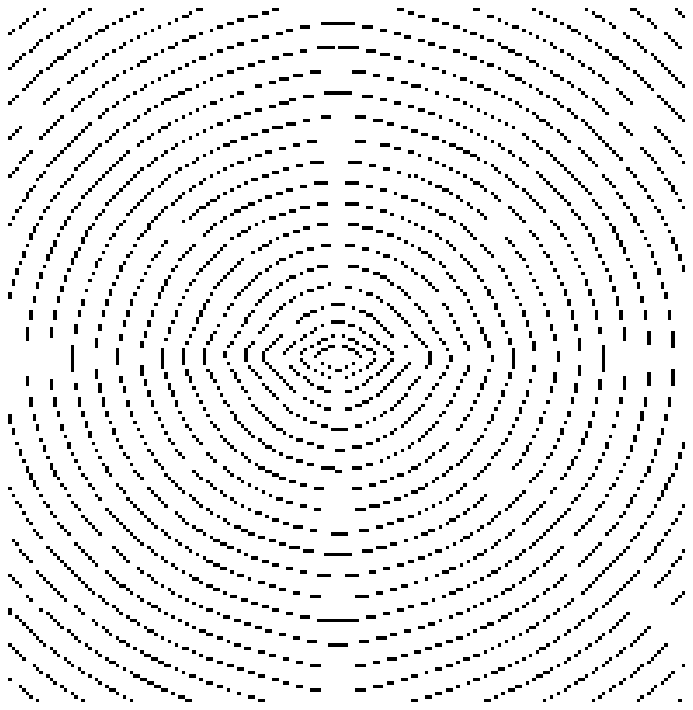}}} \hfill \=
\fbox{\resizebox{4.0cm}{4.0cm}{\includegraphics[clip=true]{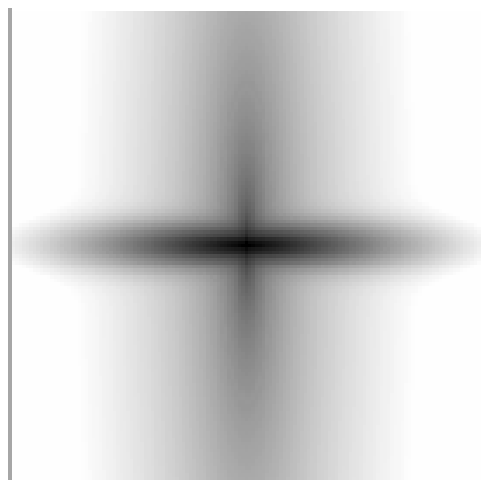}}} \= \\
\fbox{\resizebox{4.0cm}{4.0cm}{\includegraphics[clip=true]{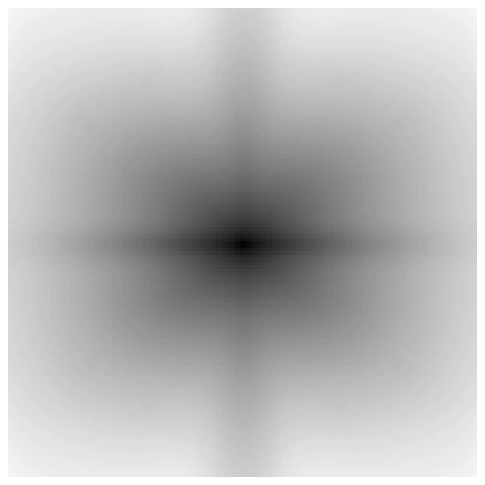}}}\>
\fbox{\resizebox{4.0cm}{4.0cm}{\includegraphics[clip=true]{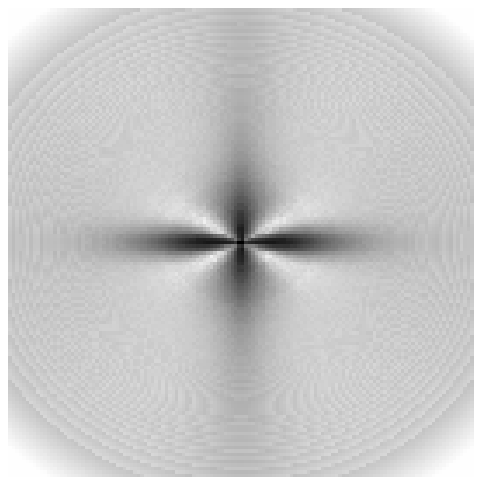}}}\> 
\fbox{\resizebox{4.0cm}{4.0cm}{\includegraphics[clip=true]{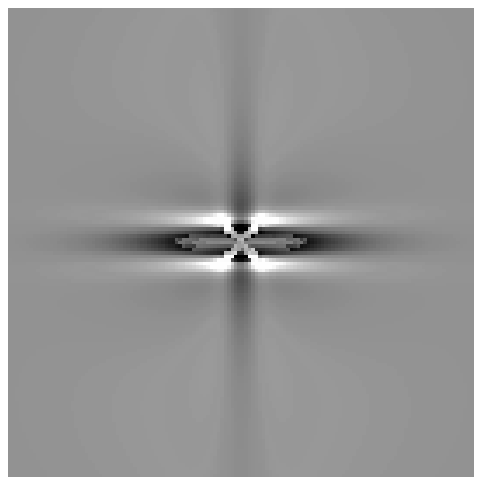}}}\> \\
\end{tabbing}
\caption{As Fig.\ref{DBNEI1}, but for model {\bf DBNEI$_1$}.
}
\label{DBNEI1}
\end{figure*}

\newpage
\clearpage

\begin{figure*}[hbpt]
\begin{tabbing}
\fbox{\resizebox{4.0cm}{4.0cm}{\includegraphics[clip=true]{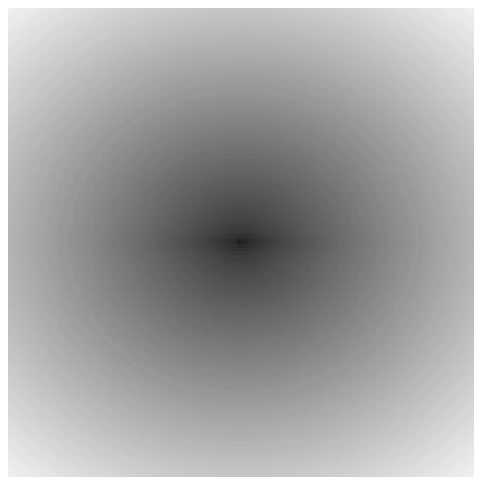}}} \hfill \=
\fbox{\resizebox{4.0cm}{4.0cm}{\includegraphics[clip=true]{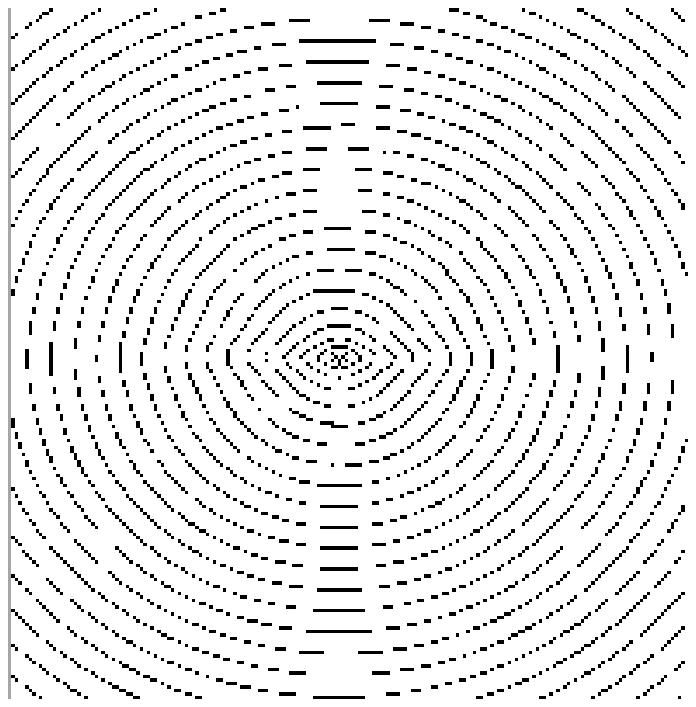}}} \hfill \= 
\fbox{\resizebox{4.0cm}{4.0cm}{\includegraphics[clip=true]{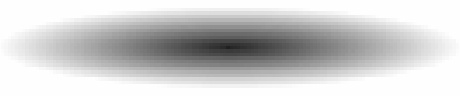}}} \= \\
\fbox{\resizebox{4.0cm}{4.0cm}{\includegraphics[clip=true]{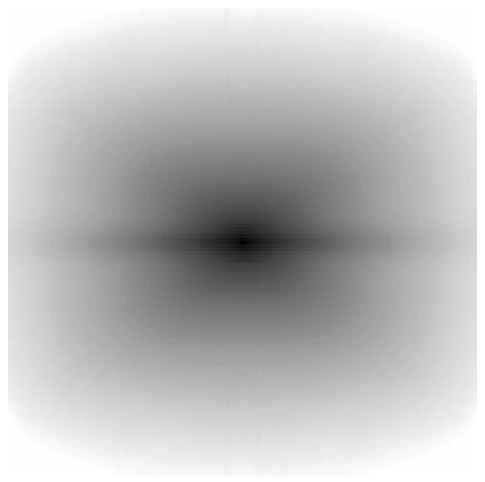}}}\>
\fbox{\resizebox{4.0cm}{4.0cm}{\includegraphics[clip=true]{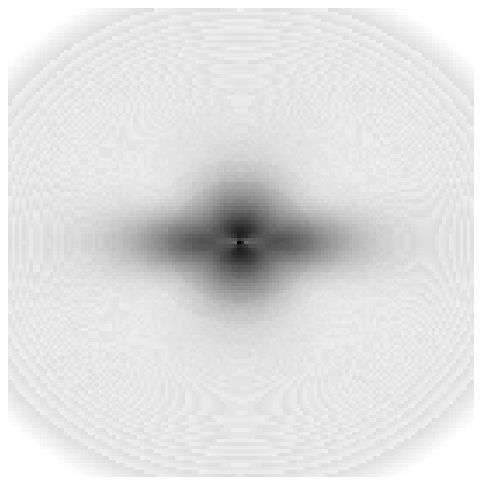}}}\> 
\fbox{\resizebox{4.0cm}{4.0cm}{\includegraphics[clip=true]{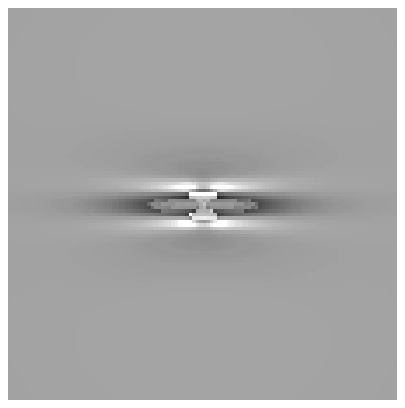}}}\> \\
\end{tabbing}
\caption{As Fig.\ref{DBNEI1}, but for model {\bf DBNI$_1$}.
}
\label{DBNI1}
\end{figure*}

\begin{figure*}[hbpt]
\begin{tabbing}
\fbox{\resizebox{4.0cm}{4.0cm}{\includegraphics[clip=true]{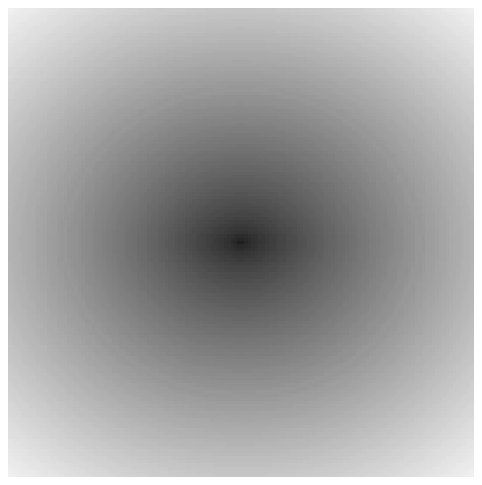}}} \hfill \=
\fbox{\resizebox{4.0cm}{4.0cm}{\includegraphics[clip=true]{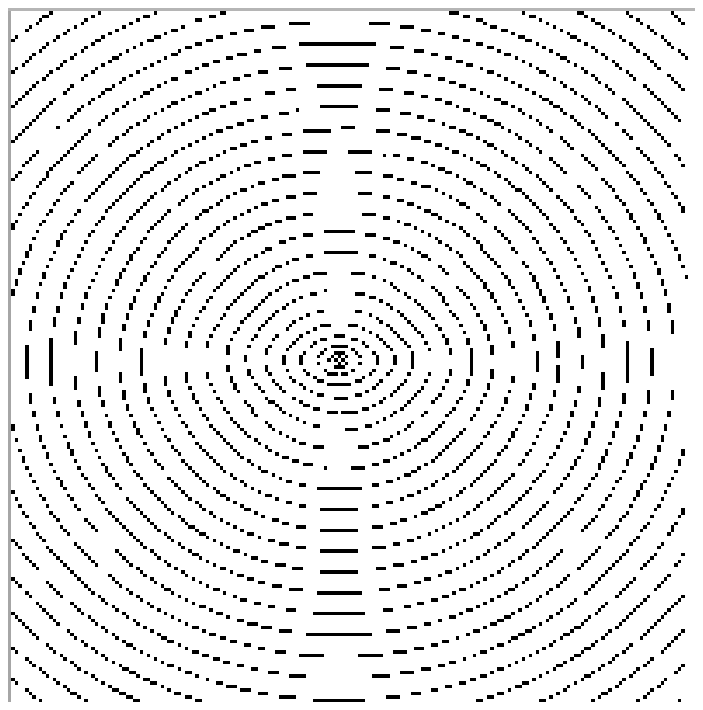}}} \hfill \=
\fbox{\resizebox{4.0cm}{4.0cm}{\includegraphics[clip=true]{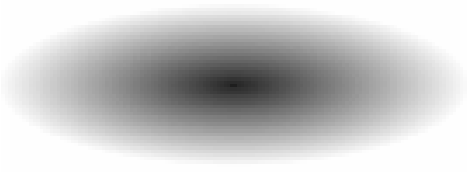}}} \= \\
\fbox{\resizebox{4.0cm}{4.0cm}{\includegraphics[clip=true]{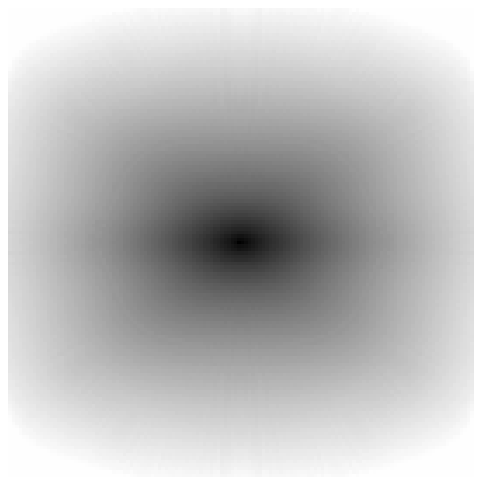}}}\>
\fbox{\resizebox{4.0cm}{4.0cm}{\includegraphics[clip=true]{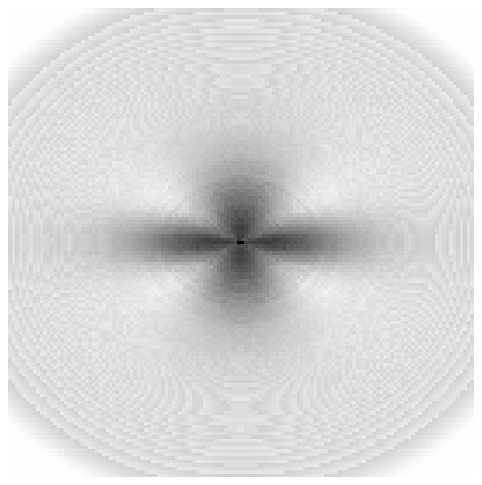}}}\> 
\fbox{\resizebox{4.0cm}{4.0cm}{\includegraphics[clip=true]{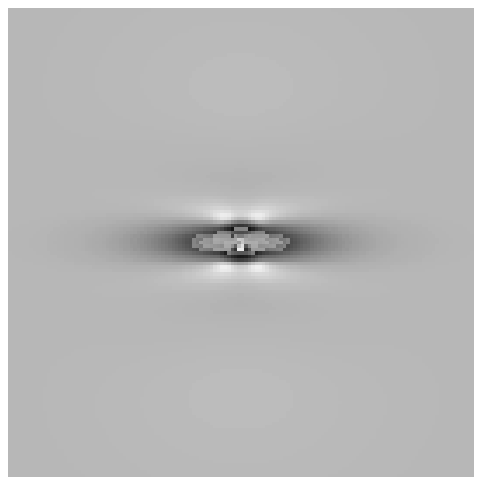}}}\> \\
\end{tabbing}
\caption{As Fig.\ref{DBNEI1} but for model {\bf DBNI$_2$}.
}
\label{DBNI2}
\end{figure*}

\newpage
\clearpage

%
\subsection{Methods to visualize the inner structure}\label{methods}
%

We apply four different techniques to visualize the substructure in the 
inner region ($\la 1$\,kpc) of NGC\,3384. The first three methods are based on the 
idea of a galaxy composed of large-scale regular components (the 
``underlying galaxy'' = disk + bulge) hosting additional, smaller-scale 
components; the methods try to separate the latter. The last method is 
based on the Laplacian pattern enhancement operator. The methods were amply 
tested by simulated images (see below).\\ 

\noindent
- {\it Profile decomposition}

The decomposition of the surface brightness profile is a conventional 
technique to characterize the basic components of a galaxy. For  NGC\,3384, 
however, the decomposition is difficult at larger scales due to the presence 
of the lense component (at $\sim 35\arcsec \ldots 170\arcsec$ along the 
major axis) and the outer disturbances (Busarello et al \cite{Busarello96}). 
At the scale of the HST images, however, the profile appears to be the composite 
of essentially three components: 
the nucleus, the IC and an approximately exponential component 
($I \propto \mathrm{exp}\,\{-r/r_0\}$, 
where $I$ is the intensity per surface unit and $r_0$ is the scalelength) which 
dominates at $a \ga 5\arcsec$ (Fig.\,\ref{profil}). We use the results from the 
ellipse fitting (MIDAS procedure FIT/ELL3 in the context SURFPHOT) to build 
a model image of this exponential component, assuming that it can be extrapolated 
from the interval $a = 6\ldots 10\arcsec$ to the innermost region. The subtraction 
of the model from the original image clearly shows the IC (Fig.\,\ref{WFPC2}), 
though the details of the residual image depend on the assumptions on the 
ellipticity of the inner isophotes of the {\it underlying} (exponential) 
component, which cannot be extrapolated with confidence. The residual image 
in Fig.\,\ref{WFPC2} is for the 
case that the ellipticity of the underlying component is identical 
with the ellipticity found from the ellipse fitting procedure.
\\

\begin{figure}[hbpt]
\includegraphics[width=5.7cm,angle=270,clip=true]{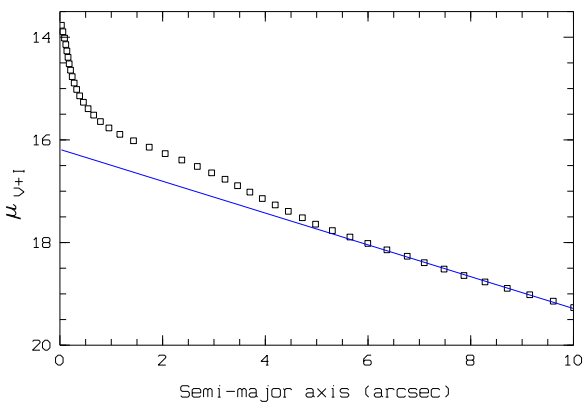}
\caption{Surface brightness of the combined {\it HST} WFPC2 F555W+F814W image 
as a function of the semi-major axis $a$ (symbols) and the linear regression for 
$a>6\arcsec$ (line).}
\label{profil}
\end{figure}

\pagebreak

\noindent
- {\it Unsharp masking}
 
This method is useful to enhance sharp features on a smooth background by 
subtracting (or dividing) a smoothed version of an image from (by) 
the original (e.g., Malin \& Carter \cite{Malin83}; Peng et al 
\cite{Peng02}). We created a blurred image by repeatedly applying a 
median filter. 
Prior to the filtering, the pixel values in the central 1\arcsec\, were 
replaced by the values of the neighbour pixels to reduce the effects 
produced by smearing the bright nucleus. Finally, the median-filtered image 
was subtracted from the original image. 
The results depend on the filter size. The residuals from the WFPC2 image are 
displayed in Fig.\,\ref{WFPC2} for a $2\arcsec\,\times 2\arcsec$ filter, those 
from the ACS images in the three filter bands are shown in Fig.\,\ref{ACS} 
for the same filter size.\\

\noindent
- {\it Ellipse fitting}

This method is also based on the idea that the underlying extended (outer) 
galaxy components can be fitted by purely elliptical iso\-photes. Hence,
additional (inner) structures of smaller scales are believed to be 
extracted by subtracting an ellipse fitting model from the original image. 
This method was often applied to analyze galaxy components; for a recent 
example see e.g. Beaton et al (\cite{Beaton06}).

However, when applied to the original image, the best fitting model is 
determined not only by the underlying components but, of course, also by the 
additional components. One solution would be masking of the expected additional structure 
and computing the ellipse model from the masked image (e.g., Lauer \cite{Lauer85}; 
Busarello et al \cite{Busarello96}). Here, we use a different approach based 
only on the analysis of the isophotes. The basic idea is that any additional 
light-emit\-ting component yields a corresponding {\it outward} distortion of the 
isophote within the angular interval covered by this component. 
(Analogously, an absorbing component, like a dust lane, produces an 
corresponding inward distortion.)
If the underlying and the additional component differ in ellipticity and/or 
position angle, the observed  isophotes can be subdivided into sections dominated 
by (A) the underlying components and (B) the underlying+additional components. 
The ellipse fitting has to be done in the A-sections only. The principle
remains the same when two additional components (here: IC and EC) are superimposed 
as is illustrated by Fig.\,\ref{isophote}. 
It is immediately clear that the subtraction of a model image from simple 
ellipse fitting yields a positive difference in the B-sections and a negative 
in the A-sections. The result is the well-known appearance of a
Maltese cross absorption feature in the residual image (see e.g. Beaton et al 
\cite{Beaton06}; their Fig. 2). The method applied here is aimed at an ellipse
model which fits to the isophotes in the A-sections only 
(IEF = ``inner ellipse fitting''): 
for each intensity level the isophote is compared with a set of fitting 
ellipses of different size. The size is the free parameter of the fit, the other
parameters are taken from the mean fitting ellipse from the FIT/ELL3 procedure.

Of course, the method can work only when the isophotes show obvious deviations 
from pure ellipses. For the inner part ($a \la 10\arcsec$) of NGC\,3384, such 
deviations are not very strong but present. 
In the inner 
$\sim 3\arcsec$, the isophotes are nearly perfect ellipses, probably 
because the IC dominates the image structure. Hence, we simply extrapolate 
the size correction from the intermediate region ($\sim 3\arcsec\ldots 6\arcsec$) 
towards the centre. The final corrected IEF image is displayed 
in panel (e) of Fig.\,\ref{WFPC2}. 
We note that the simple ellipse fitting (not shown here) of NGC\,3384 
yields a very pronounced 
cross-like feature both in emission and absorption. This feature
is considerably reduced by the IEF method and the subsequent size correction
but does not disappear completely.\\

\noindent
- {\it Adaptive Laplace filtering}
   
Finally, we processed the WFPC2 image through an ad\-aptive Laplacian filter.
Laplace filtering uses the second de\-revatives of the image and is therefore 
a useful tool of structure recognition; adaptive filtering means to look 
for the spatial frequency band containing the relevant signal and to enhance particularly 
the structure in this band (e.g. Richter et al \cite{Richter91}).
Adaptive Laplacian filtering has been successfully applied by Busarello et al 
(\cite{Busarello96}) and has led to the discovery of the IC. 
Here we applied the MIDAS procedure FILTER/ADAPTIVE with the Laplace filter 
where the parameters were chosen after extensive tests (pyramidal impulse 
response with a maximum size of 95, a significance threshold of $k = 3$, and 
an assumed Poisson noise). The result is shown in panel (f) of Fig.\,\ref{WFPC2}.
We performed extensive modelling to understand the structures in this image
(see below). 

%
\subsection{Tests with simulated galaxies}\label{tests}
%

Here we describe synthesized images from simulated 
galaxies which were analyzed in the same way as the {\it HST} images
of NGC\,3384. The aim is to provide a better 
understanding for the structures revealed and/or produced by the 
image analysis techniques. It should be explicitely pointed 
out that it is not the intention of this section to search for 
a kind of ``best fit'' models.  
 
In general, the models consist of five components: (a) the extended, inclined galaxy disk, 
(b) a bulge, (c) a compact central component, called ``nucleus'', 
(d) an inner component, called IC, which is elongated in the same direction as 
the projected disk, and (e) an elongated inner component, called EC, stretching 
perpendicular to the IC. 
The model images were constructed using 
the MIDAS procedure CREATE/IMAGE with an usual exponential law for a thin, 
face-on galaxy disk, an $r^{1/4}$-law for an spherical-symmetrical bulge, 
and a Gaussian with $\sigma_{\rm x,y} = 1$\,px for an unresolved nucleus. 
For the IC we adopt an inclined exponential disk at $PA=90\degr$ with a scale 
length between 0.02 and 0.1 of that of the extended galaxy disk.
The EC is represented by an extended, elongated component at $PA = 0\degr$ with a 
surface brightness 
distribution following an $r^{1/4}$-law, though other luminosity profiles were 
simulated as well. The inclination of the IC is varied between 
$i = 80\degr$ and $60\degr$ (less inclined disks become photometrically 
undetectable; Rix \& White \cite{Rix90}) where $i$ is the angle between the 
line of sight and the direction perpendicular to the plane of the IC. All 
components are co-centric.
For four realizations the synthesized images are displayed in 
Figs.\,\ref{DBN} to \ref{DBNI2} along with the results from the image
processing. The intensity levels in the contour plots
are equal-spaced with steps of $\Delta \log\,I = 0.1$.
The models are listed in Table\,\ref{modelsim}.
Disk, bulge, and nucleus have the same parameters in all displayed images. 
Model DBN is used only to check for numerical artifacts from the 
image analysis.  The comparison of Figs.\,\ref{DBNEI1} and \ref{DBNI1} allows
to evaluate to effect of an EC-like component and the comparison 
of Fig.\,\ref{DBNI1} with Fig.\,\ref{DBNI2} illustrates the effect of 
changing the inclination of the IC. The extracted structures from the image 
analysis can be compared directly with the images of the corresponding input 
components. The additional components are clearly recognized in the 
results from all methods, but it is also obvious that all methods produce 
deformations or ``artifacts''.  

\begin{table}[hbpt]
\caption{Simulated galaxies from Fig.s\,\ref{DBN} to \ref{DBNI2}}
\begin{tabular}{ll}
\toprule
model                    &  components \\
\midrule
{\bf DBN}                & disk, bulge, nucleus \\
{\bf DBNEI$_1$}          & disk, bulge, nucleus, EC, IC\,($i=80\degr$) \\
{\bf DBNI$_1$}           & disk, bulge, nucleus, IC\,($i=80\degr$) \\
{\bf DBNI$_2$}           & disk, bulge, nucleus, IC\,($i=70\degr$) \\
\bottomrule
\end{tabular}
\label{modelsim}
\end{table}
 
\begin{figure}[hbpt]
\begin{tabbing}
\fbox{\resizebox{3.8cm}{3.8cm}{\includegraphics[clip=true]{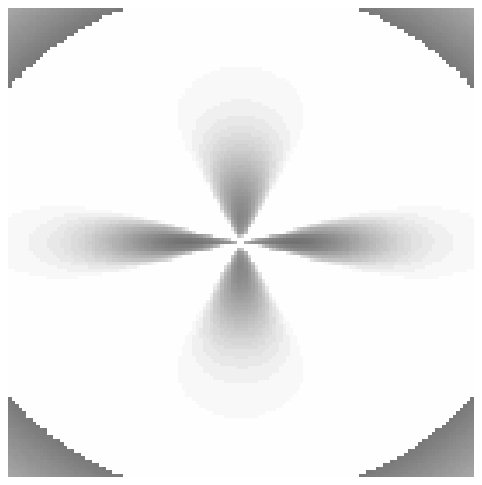}}} \hfill \=
\fbox{\resizebox{3.8cm}{3.8cm}{\includegraphics[clip=true]{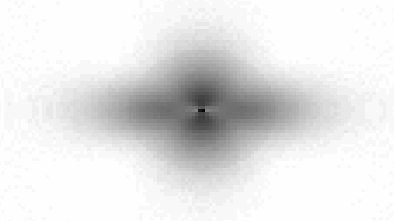}}} \= \\
\end{tabbing}
\caption{Comparison of the residual image from the simple ellipse fitting (left) and the
corrected IEF (right) for the same model (DBNI$_1$).}
\label{comp_simple_eifc}
\end{figure}

The unsharp masking provides a good impression of the basic 
structure of the components, but the underlying components are still 
very prominent. 
The residuals from the subtraction of corrected IEF models  
are contaminated by effects due to the discrete step size of the intensity 
of the fitting ellipses especially in the outer regions. In the inner region, 
the well-known Maltese cross is produced
(Fig.\,\ref{DBNI1}). Compared to the residual image from the simple ellipse
fitting, the Maltese feature is substantially suppressed in 
the corrected IEF image (Fig.\,\ref{comp_simple_eifc}). However, it does never disappear completely,
even in the model galaxies without EC (Figs.\,\ref{DBNI1} and \ref{DBNI2}).
The reason is that the method works only as long as the isophotes clearly 
deviate from pure ellipses, as mentioned already in Sect.\,4.2. For 
small deviations, the A- and B-sections (Fig.\,\ref{isophote}; Sect.\,4.2)
are not precisely enough defined. This is the case when the 
additional components have a similar shape and orientation as the underlying 
components, as is illustrated by the comparison of Figs.\,\ref{DBNI1} and 
\ref{DBNI2}: the galaxy models are the same with the only difference in the 
inclination of the IC. Because of the lower inclination in Fig.\,\ref{DBNI2}, 
the underlying component and the additional component (IC) show less
significant differences and the isophotes of the model galaxy hence
show smaller deviations from pure ellipses. Due to the assumed symmetry, 
small systematical errors in the determination of the  A- and B-sections produce
cross-like structures. This can result also, of course, in apparent 
absorption features, even though the method of the {\it inner} ellipse fitting
was constructed to avoid such artificial structures. 

Finally, the Laplace filtering yields noticeable signals at positions where 
the intensity gradient abruptly changes. The filtered images allow 
to estimate the structure and also the size of the inner components.
In particular, there are strong negative signals at either side as well of the 
IC as of the nucleus parallel to IC's major axis. At somewhat larger 
distance from the centre, two extended side-lobes appear in the direction 
of the EC in models where there is no EC. Theses features are very faint
and are not clearly recognizable 
on the reproductions in Figs.\,\ref{DBNI1} and \ref{DBNI2}

%
\subsection{Results}\label{results} 
%

The results from the combined WFPC2 image are summarized in Fig.\,\ref{WFPC2}.
The IC is clearly indicated in the final images from all four image processing 
techniques. The residual images from the profile decomposition and from the 
unsharp masking (both are reproduced with high contrast) reveal an ellipsoidal 
structure at P.A.$=46\degr$ with a maximum 
semi-major axis of $\sim 5\ldots6\arcsec$. This is consistent with the structures seen in the images 
from the ellipse fitting procedure and the Laplace filtering 
(compare Figs.\,\ref{WFPC2} and \ref{DBNI2}). The Laplace filtered image 
indicates, in addition, a minimum in the radial change of the slope of the 
radial profile along the major axis at about $2\arcsec$ from the centre, 
indicating that the profile of the IC becomes shallower towards the centre.
This characteristics becomes directly visible in Fig.\,\ref{ell_from_profil_decomp} 
where the surface brightness profile of the residual component is shown after 
subtraction of the model from the profile decomposition. Obviously, the radial intensity 
profile of the IC is not adequately described by the usual exponential law of stellar 
galactic (outer) disks. Below $\sim 1\arcsec$, the surface brightness increases 
rapidly towards the centre. 

On the unprocessed {\it HST} images, the maximum ellipticity of the IC is 
$\epsilon_{\mathrm{ max, IC}} = 0.44$ (Fig.\,\ref{Nuker}). Busarello et al. (their Fig.\,4) 
give smaller values of $\sim 0.4$ for the NTT image and $\sim 0.3$ for the 
other images. The corresponding value from our Calar Alto data is $\sim 0.35$,
in good agreement with Busarello et al.  Because these differences are 
obviously a resolution effect, the best value is that from the {\it HST} observations. 
The measured ellipticity is expected to be affected also by the profile of 
the underlying galaxy. After subtraction of the model from the profile decomposition 
we measure indeed a higher value of $\epsilon_{\mathrm{ max, IC}} = 0.50$
(Fig.\,\ref{ell_from_profil_decomp}). If the ellipticity is interpreted as due to
an inclined disk, the IC has nearly the same inclination as the outer disk 
($\epsilon_{50} = 0.49$; see Sect.\,\ref{UBVR}).
The position angles, however, are slightly different: $46\degr$ for the IC and
$51\degr$ for the outer disk.

\begin{figure}[hbpt]
\includegraphics[width=6.2cm,angle=270,clip=true]{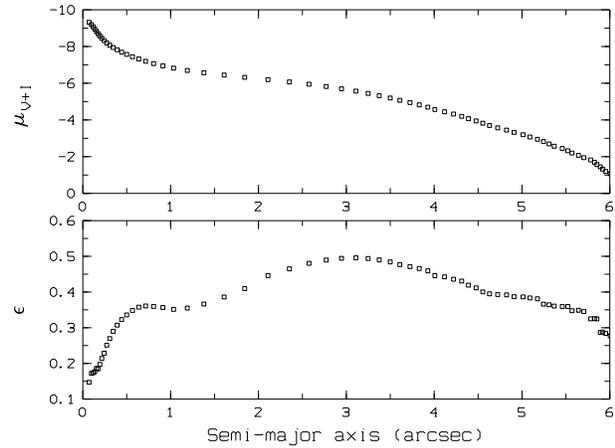}
\caption{Surface brightness $\mu$ and ellipticity $\epsilon$ of the residual WFPC2 
image after subtraction of the model image from profil decomposition.}
\label{ell_from_profil_decomp}
\end{figure}

The EC is weakly indicated in all images, also in the Laplace filtered image 
(though it is not visible in the reproduction in Fig.\,\ref{WFPC2} where 
the intensity cuts were chosen to visualize the structure of the IC). 
Our images corroborate the earlier finding 
(Busarello et al \cite{Busarello96}; Sil'chenko 
et al \cite{Silchenko03}) that the elongation of the EC is nearly perpendicular 
to the IC. There is no indication for inner structure in the EC.
However, we find that the contribution of the EC to the total intensity 
per surface unit does not exceed 8\%, while Busarello et al report a 
maximum of 15\% at larger distances from the centre. This is in agreement
with the visual impression that the relative strength of the EC has its
maximum at $\sim 15\arcsec$ from the centre and is much fainter inside. 
We notice that such a behaviour is not reproduced by the bars in our
simulations.

The residuals from the subtraction of the median filtered images
in three filter bands of the {\it HST} ACS images are shown in 
Fig.\ref{ACS} along with the corresponding contour plots. 
The intensity levels in the contour plots are equal-spaced with steps 
of $\Delta \log\,I = 0.1$.
At longer wavelengths (F555W; see also the 
combined image in Fig.\,\ref{WFPC2}), the IC is traced out to 
distances of about  4\farcs5$\ldots$5\farcs5 (about 250 $\ldots$ 300\,pc).
From the comparison of the residual images from the profile decomposition 
with the original images, we estimate that the surface intensity of the 
IC on the major axis at $a \approx 3\arcsec$ is probably as high as 
30 to 40 per cent
of the galaxy (see also Figs.\,\ref{Nuker} and \ref{profil}). This is 
clearly at variance with the upper limit of 5\% estimated by  Busarello 
et al. The most likely reason for this discrepancy is the higher
resolution of the {\it HST} images.  
 
The centre of NGC\,3384, and also the IC, are clearly detected at 2500\AA\, 
(Fig.\,\ref{ACS}). This UV background appears to be caused mainly by low-mass, 
helium-burning stars in the horizontal branch and subsequent phases of 
evolution (see O'Connell \cite{O'Connell99}). An important
consequence is that the UV background allows to detect efficiently faint dust 
extinction features (see Sect.\,6).

\begin{figure}[hbpt]
\begin{tabbing}
\fbox{\resizebox{3.8cm}{3.8cm}{\includegraphics[clip=true]{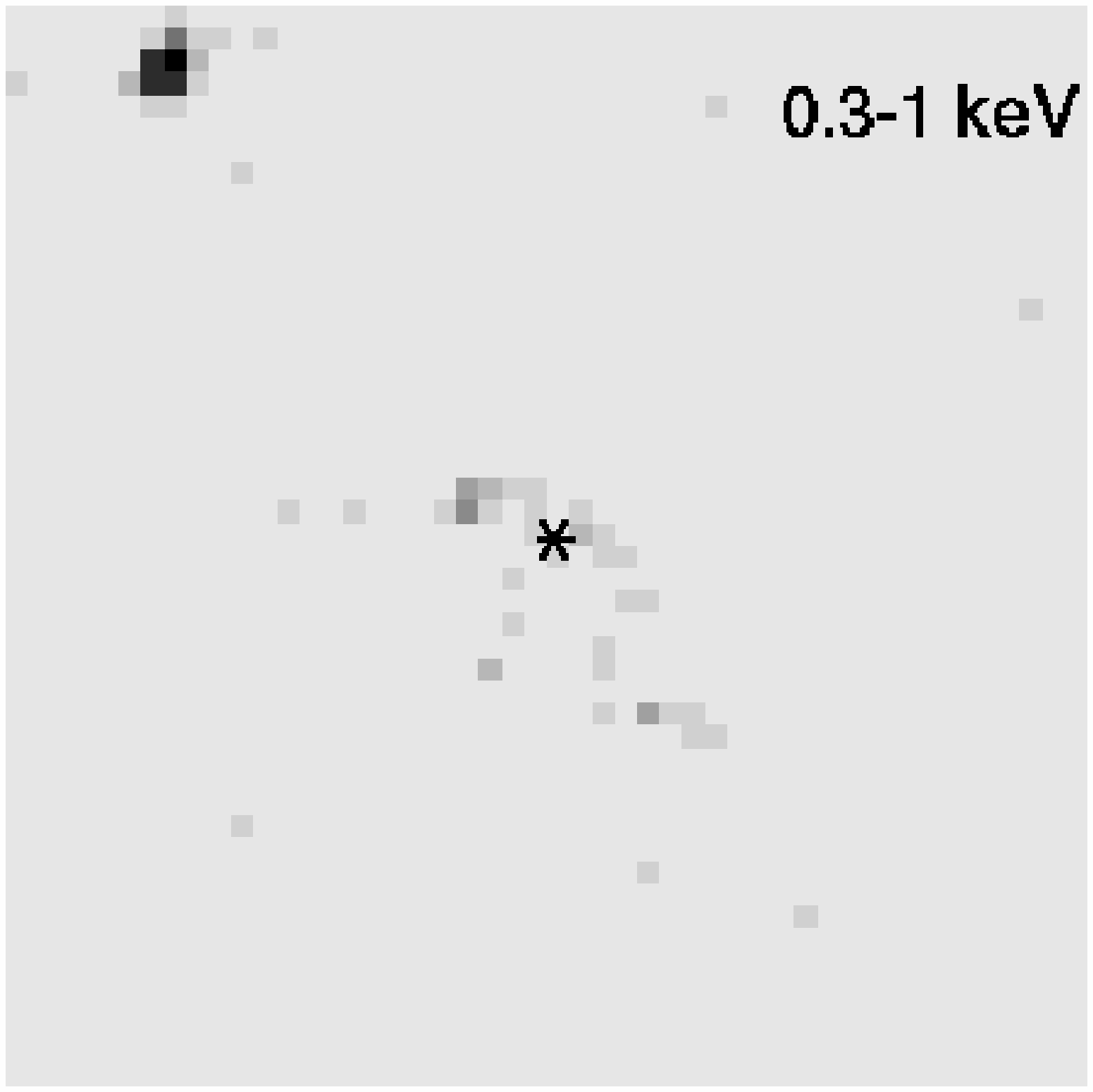}}} \hfill \=
\fbox{\resizebox{3.8cm}{3.8cm}{\includegraphics[clip=true]{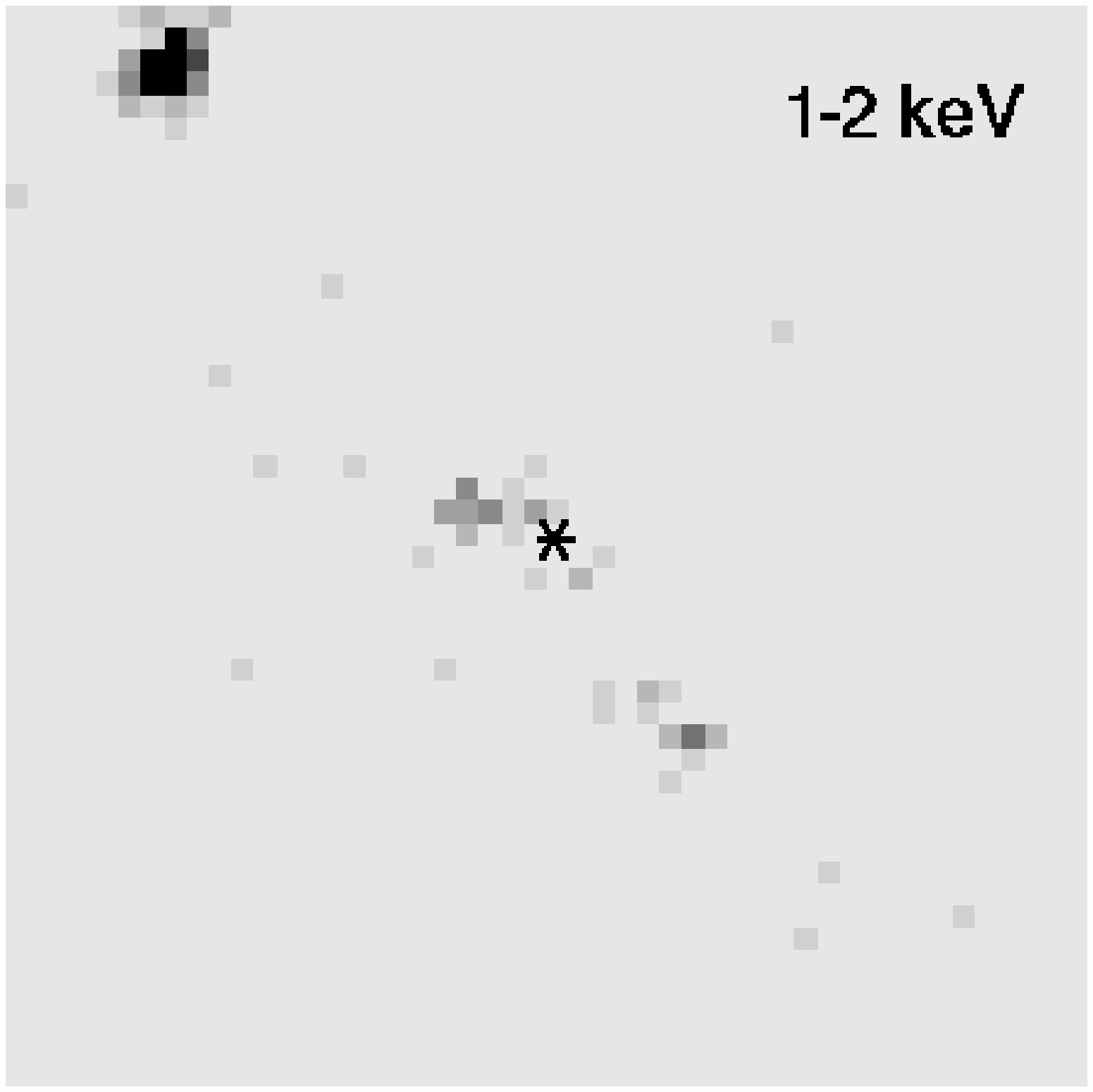}}} \= \\
\fbox{\resizebox{3.8cm}{3.8cm}{\includegraphics[clip=true]{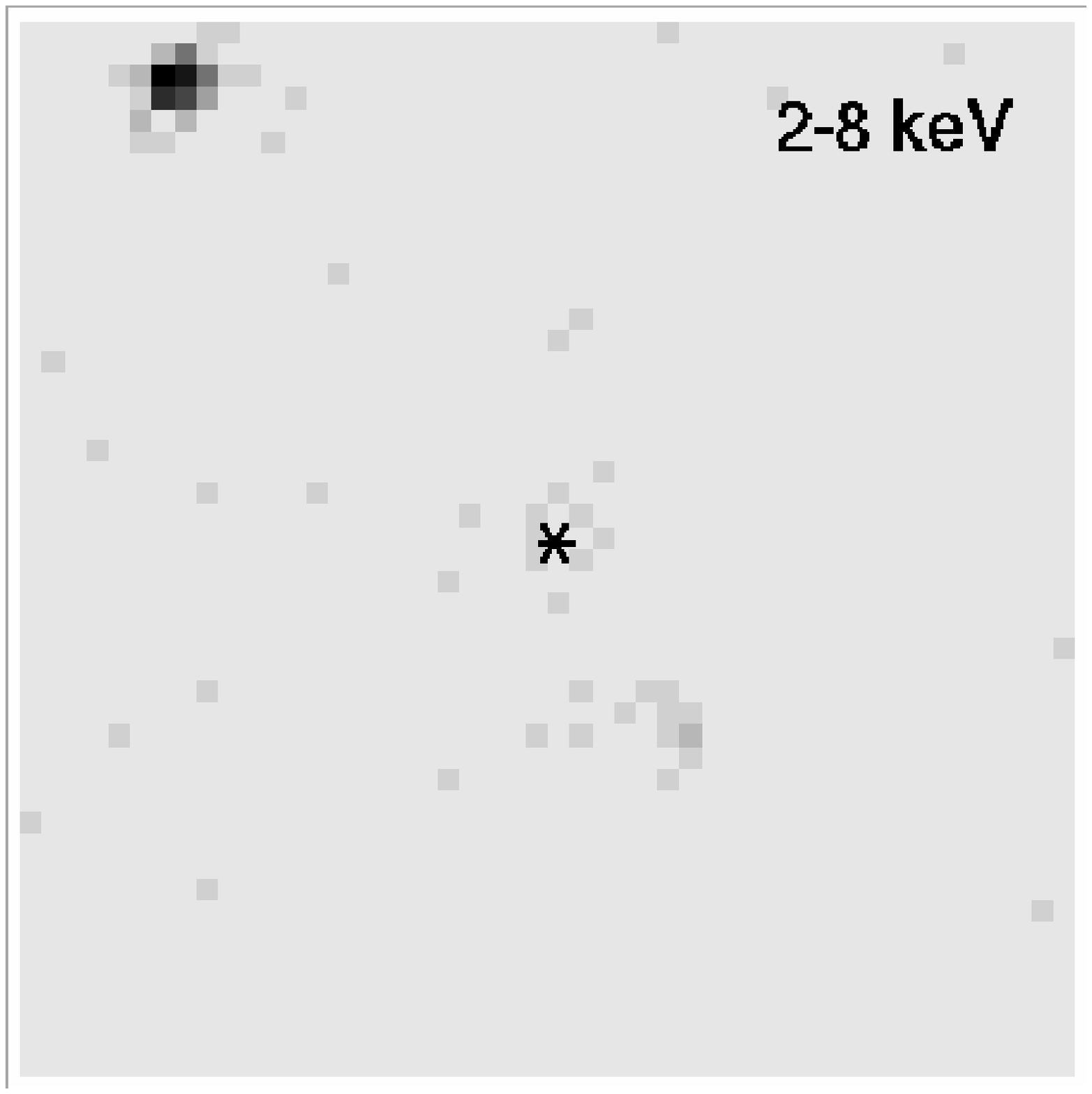}}} \hfill \=
\fbox{\resizebox{3.8cm}{3.8cm}{\includegraphics[clip=true]{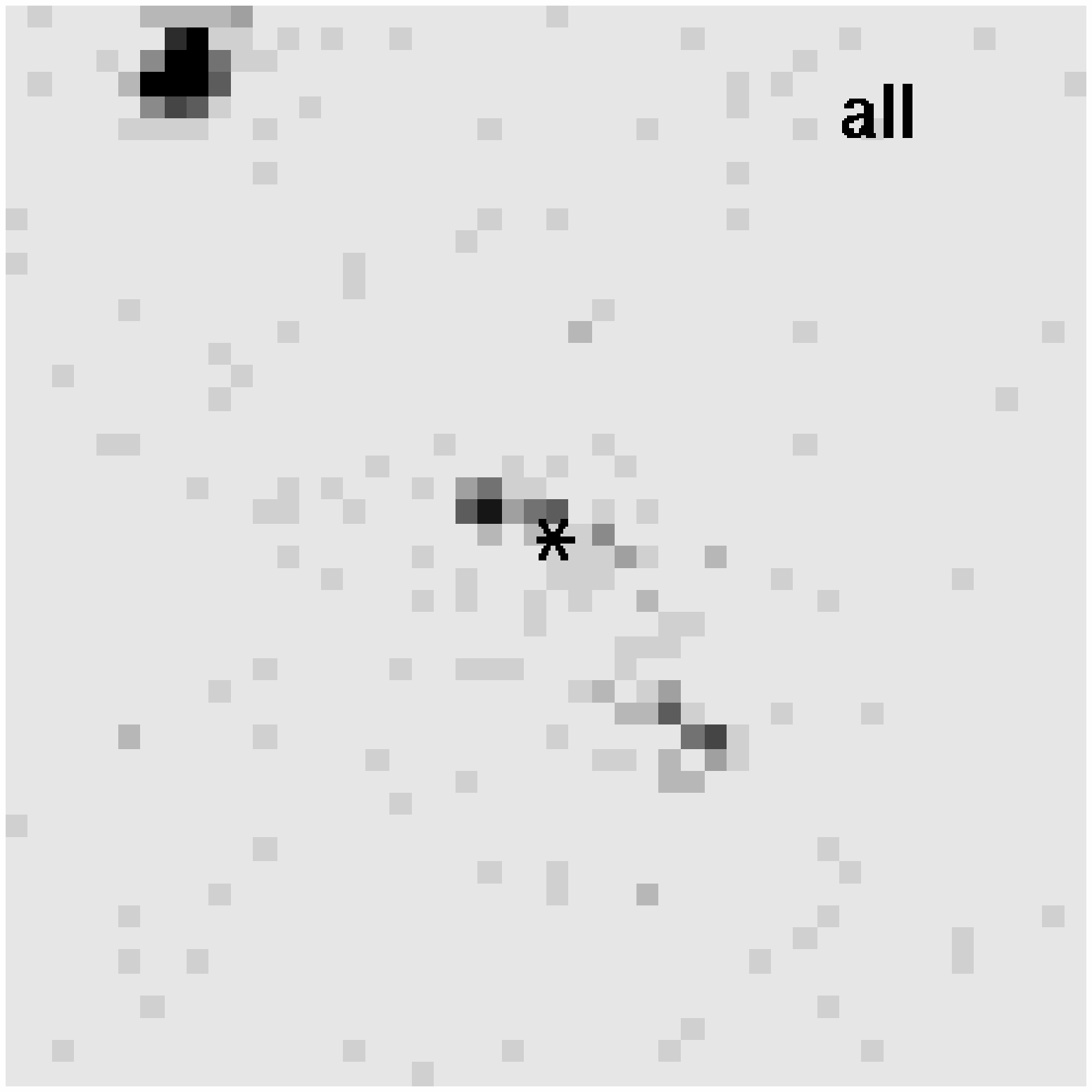}}} \= \\
\end{tabbing}
\caption{{\it CHANDRA} image of the inner 24\arcsec of NGC\,3384 (same scale 
as in Fig.\,\ref{WFPC2}). The position of the optical centre of the galaxy 
is marked by the black asterisk in the image centre.}
\label{chandra}
\end{figure}

Finally, the {\it CHANDRA} image of NGC\,3384 is worth noticing 
(Fig.\,\ref{chandra}). There are three, perhaps four or five, 
fain\-ter X-ray sources in the image and a brighter source at  the 
NE corner. One source is close to the centre of the galaxy, though 
presumably not identical with the nucleus. The position of the optical 
centre was determined by five 
X-ray sources in the whole {\it CHANDRA} field which were unambiguously 
identified with optical sources.
NGC\,3384 is ``X-ray faint'' in the sense that it is devoid of a large 
gaseous X-ray halo. In X-ray faint early-type galaxies, low-mass 
X-ray binaries (LMXBs) 
are known to account for a large fraction of the total X-ray emission 
(Fabbiano \cite{Fabbiano06}). Hence, it is tempting to speculate that the 
sources seen in Fig.\,\ref{chandra} are mostly LMXBs; a more detailed 
discussion is beyond the scope of this paper. Here, we just notice 
that the distribution of the X-ray sources shows approximately the same 
orientation as the IC.

%
\section{Age estimation}   
%

An excellent analysis of the ages of the stellar population in the centres 
of the three brightest Leo\,I galaxies, NGC\,3368, NGC\,3379, and 
NGC\,3384, has been presented by 
Sil'chen\-ko et al \cite{Silchenko03}. Based upon Lick indices from 
integral field spectroscopy, they discovered separate circumnuclear 
subsystems in all three galaxies. In the centre of NGC\,3384, a chemically 
distinct subsystem was found with an age of about 3\,Gyr, a
super-solar metallicity $Z \approx 0.04$\, ([m/H] = 0.35) and 
an overabundance of $\alpha$ elements ([Mg/H] $\approx 0.3$). The extent 
of this component was found to be comparable to or smaller then the resolution limit 
of $\sim 1\arcsec$. In the nearest vicinity of the nucleus, the 
metallicity and the relative abundances drop to solar values while the 
mean age increases to 7-8\,Gyr. The use of the spectroscopic Lick indices
provides the advantage of being widely unaffected by dust. 
On the other hand, the limited field of view of integral field spectrographs
can be a drawback sometimes. For the investigation of the stellar 
population on larger scales in nearby galaxies, the multi-colour photometry 
hence remains an important tool. 

\begin{figure}[htbp]
\includegraphics[width=6.0cm,angle=270,clip=true]{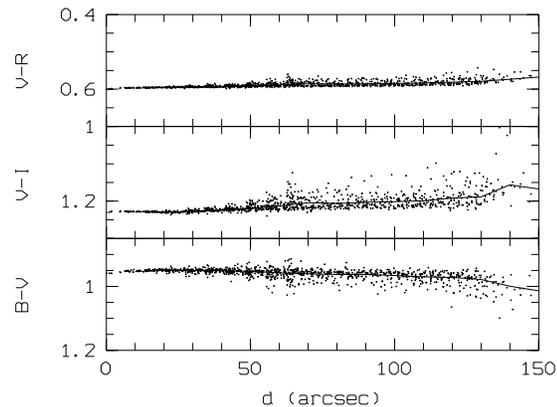}
\caption{The colours for points located along or parallel to the major axis
versus distance $d$ from the centre. The solid lines are running means.}
\label{ciscan}
\end{figure}

\begin{figure}[htbp]  
\includegraphics[width=5.7cm,angle=270,clip=true]{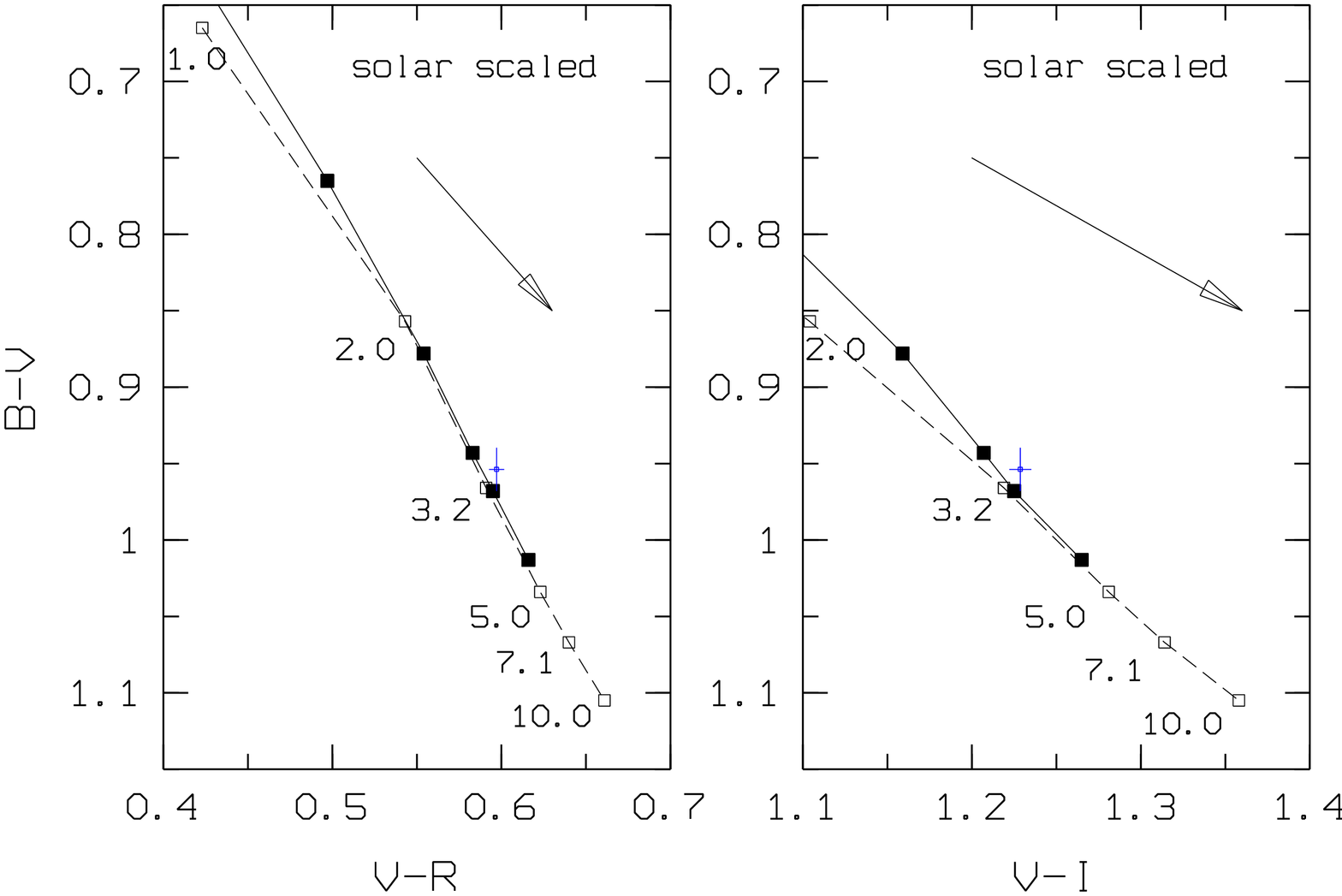}
\includegraphics[width=5.7cm,angle=270,clip=true]{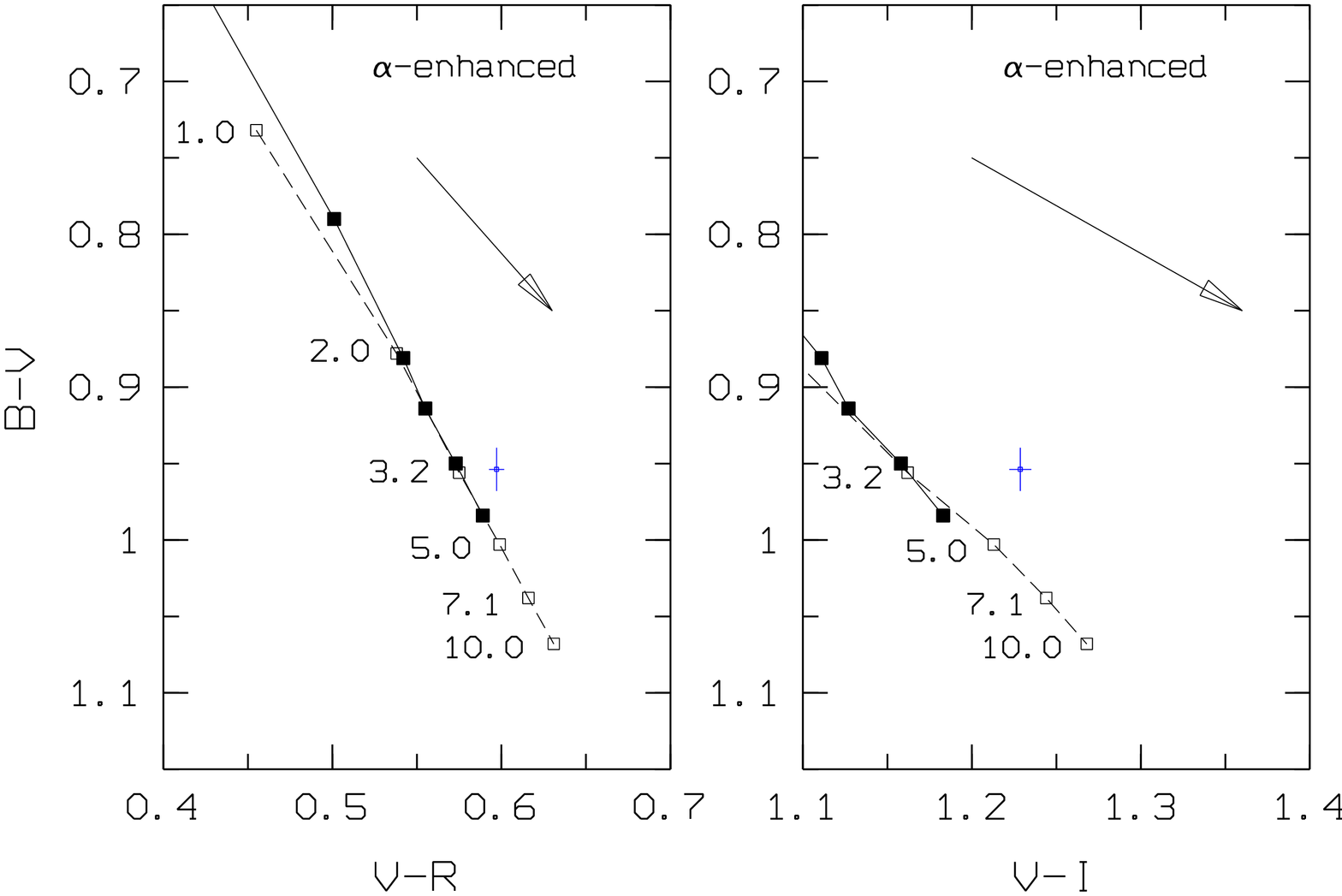}
\caption{Colour-colour diagrams with evolutionary tracks of single 
starburst populations with solar-scaled (top) and $\alpha$-enhanced (bottom) 
relative abundances from Salasnich et al (\cite{Salasnich00}) 
for metallicity $Z = 0.019$ 
(solid line, filled squares) and $Z = 0.04$ (dashed line, open squares),
respectively. The points along the tracks correspond to ages of 
10.0,7.1,5.0,3.2,2.0, and 1.0\,Gyr (bottom to top; the youngest model points 
are beyond the boundaries for some tracks; for lucidity only the age points 
on the 
super-solar metallicity tracks are labelled). The mean colours of the stellar 
population in NGC\,3384 are marked by the small dot with $1\sigma$ error bars. The arrows are
the reddening vectors for standard Galactic dust and $E(B-V)=0.1$\,mag.
}
\label{ccd}
\end{figure}

Here we use our multi-colour photometry on the Calar Alto images to quantify
characteristics of the large-scale distribution of stellar populations in 
NGC\,3384 by the comparison with model colours. Colour indices were measured 
in  scans parallel to the major axis of the galaxy and parallel its the 
minor axis, respectively. Measurement apertures of 1\arcsec to $2\farcs5$ diameter were 
used separated by $2\farcs5$. The measurements for the scans parallel to the 
major axis are shown in Fig.\,\ref{ciscan}, the results for the scans parallel 
to the minor axis are very similar. We conclude that the colours do not 
significantly vary with position on large scales, in agreement with Busarello 
et al. In particular, there are no significant systematic large-scale colour 
gradients. The mean values ($\pm 1\sigma$ uncertainties) of the colour
indices are 
$(B-V, V-R, V-I) = (0.954\pm0.014, 0.597\pm0.045, 1.229\pm0.065)$.
In Fig.\,\ref{ccd}, we compare the mean colour indices 
with the synthetical stellar population models from 
Salasnich et al (\cite{Salasnich00}). We conclude that the dominant stellar 
population in NGC\,3384 has an age of about 5 to 7\,Gyr and approximately 
solar metallicity, in good agreement with the results from Sil'chenko et al.

It was actually one of the primary aims of our multi-colour photometry to 
look for large-scale variations of the dominant stellar population.\,
Expected colour\, differences can be estimated by the comparison of 
the results from Sil'\-chenko et al for the nucleus and the host galaxy
with the population synthesis models from Salasnich et al.
In the terminology of the models, the nucleus corresponds 
to 3\,Gyr, $Z=0.04$, and $\alpha$-enhanced abundances, the main body of the 
galaxy corresponds to 7\,Gyr, $Z=0.019$, and solar-scaled abundances.
The colour differences of the two corresponding model points are \, 
$(\Delta(B-V), \Delta(V-R), \Delta(R-I)) = (0.012, 0.020, 0.063)$, i.e. very 
small. 

Let us now assume that the EC and the IC are made up of the same stellar 
population as the nucleus. In that case, the colour difference between the
host galaxy and the areas covered by these components is even 
smaller because they contribute only a fraction to the total 
light: $\le 15$\% for the EC (Busarello et al \cite{Busarello96}) and 
$\la 40$\% for the IC (Sect.\,4). From these values 
one easily estimates a colour difference relative to the underlying galaxy of 
$\Delta(B-V) < 0.002$\,mag for the EC and less than $0.005$\,mag for the IC, 
too small to be clearly detectable. Hence, our photometry 
does not allow to decide whether the formation of the EC and IC 
is related to the same event in the evolution of NGC\,3384 which triggered 
the starburst in the nucleus.   
  
We explicitely note that it was not our intention to measure colour 
gradients very close to the centre. The positions of the measurement apertures 
were not selected with special consideration of colour changes within 
the innermost few arcseconds. Systematic reddening in the nearest
vicinity of the nucleus has been reported already in several studies 
done with higher resolution
(Busarello et al \cite{Busarello96}; Sil'chenko et al \cite{Silchenko03}; 
Lauer et al. \cite{Lauer05}). Some of these colour values have to be taken 
with care since the images were not accurately calibrated 
(including in particular $U-V$ in Fig.\,\ref{Nuker} of the present paper,
but also $V-I$ in Lauer et al, and the V magnitudes in Busarello et al).  
The $B-V$ colour profiles shown by Busarello et al indicate a 
colour difference of  $B-V \approx 0.10\ldots0.15$\,mag between their 
innermost point at $a = 1\arcsec$ and the region at $\ga 2\arcsec$ 
along the minor axis and $\ga 5\arcsec$ along the major axis, respectively. 
Along the major axis, most of the reddening appears to be within 
the innermost $\sim 2\arcsec$. The higher resolution of the {\it HST} images 
indicates that the reddening is probably even stronger concentrated
toward the nucleus.

%
\section{Discussion: the dust in the central region of NGC 3384}
%

With the ages and metallicities from the spectroscopic Lick indices for
(a) the nucleus and (b) the rest of the galaxy, on the one hand, and 
the good agreement of these data with our mean colours, on the
other hand, the colour gradients close to the nucleus remain to be explained. 
From Fig.\,\ref{ccd}, it is immediately clear that a reddening 
of about $E(B-V) \approx 0.1\ldots0.2$\,mag toward the nucleus cannot be 
due to the change in 
the stellar population. A reasonable assumption is reddening due to dust. 
The issue of dust in the centre of NGC\,3384 will be discussed 
in detail below.

The possibility that dust lanes are common to early type galaxies 
was discussed already by Lauer (\cite{Lauer85}).
The high-resolution of the {\it HST} images allows to detect small-scale 
dust extinction features. It is well known from such studies that dust 
features are common in the central regions of early-type galaxies.
In the sample of nearby early-type galaxies studied by Lauer et al 
(\cite{Lauer05}), dust is visible in 47\% of the 177 galaxies. No
dust obscuration was found in NGC\,3384 on the WFPC2 images 
(Tomita et al \cite{Tomita00}; Lauer et al \cite{Lauer05}). 
The surface brightness distribution appears smooth also on the unsharpe 
masked ACS F555 and F330 images in Fig.\,\ref{ACS}. However, dust absorption 
features are detectable best in the UV due to the high selective UV extinction. 
The ACS F250W image indicates some faint fuz\-zy structures, 
most discernible on the contour plot. If these structures are due to dust, 
e.g. a faint dust lane related to the EC, the {\it maximum} local 
extinction is estimated to $A_{\lambda2500} \sim 0.5$\,mag, corresponding to 
$E(B-V) \le 0.07$\,mag for standard Galactic dust. The colour index 
$U-V$ (Fig.\ref{Nuker}) seems to be slightly ($\sim 0.05$\,mag) enhanced 
in the middle of the IC ($a \approx 2\ldots3\arcsec$) and is
increasing towards the nucleus at $a \la 1\arcsec$. 

Can we exclude additional reddening due to a smoothly distributed 
dust component? In order to answer this question, we estimate the dust extinction 
from the amount of cool gas. Early-type galaxies are known to have only a 
low fraction of 
mass in cool interstellar matter (ISM). Moreover, the measured gas masses in 
S0 galaxies are even less than 10\% of the masses expected to be 
returned to the ISM by evolved stars over a Hubble-time, $t_{\rm H}$. 
Gas masses of a sample of 27 nearby S0 galaxies, including NGC\,3384, were
derived by Sage \& Welch (\cite{Sage06}). With a total mass of 
5\,10$^6 \mathrm{m}_\odot$ 
for the cool ISM (neutral and molecular gas), NGC\,3384 is among the most 
gas-poor galaxies in this sample. Nevertheless, dust can have an effect if 
it is strongly concentrated.  
Many galaxies have peak CO emission in the centre (Regan et al \cite{Regan01}; 
Helfer et al \cite{Helfer03}) and molecular gas is known to be more centrally 
concentrated in galaxies with bars (Seth et al \cite{Seth05}). 
For the most gas-poor S0 galaxies studied by Sage \& Welch, the cool gas 
is dominated by the molecular phase which seems to be concentrated towards 
the galaxy centres while the neutral gas is probably more widely distributed 
(Sage \& Welch \cite{Sage06}). A median size of the CO-emitting region of 
0.7\,kpc has been derived by Welch \& Sage (\cite{Welch03}). This corresponds 
to $12\arcsec$ for NGC\,3384, if we assume the distance of 8.1 Mpc used by these 
authors, and scales to 1.0\,kpc for the distance used here. 
From the observed gas mass in this volume, we estimate the effect of dust 
assuming that 
(1.) the ISM in the central part of NGC\,3384 is completely dominated by 
the molecular gas, 
(2.) the molecular gas is located completely within the inner 1\,kpc,
(3.) the dust-to-gas mass ratio corresponds to the standard value of 0.01, 
(4.) the dust density is independent of position, and
(5.) the dust acts as an extinction screen. 
For the standard Galactic extinction curve we find
$A_{\rm \, V} \approx 0.3$\,mag and $E(B-V) \approx 0.1$\,mag for the centre of 
NGC\,3384. 
  
As a consistency check, we compare the estimated dust mass with the IRAS 
data. From the IRAS SCAN\-PI data we find a 12\,$\mu$m flux density of 
$f_{\, 12} = 0.13$\,Jy, but no signal at longer wavelengths. This does not 
fit the spectral
energy distribution of dust at a typical temperature of $T_D \approx 30$\,K 
(assuming the usual $\lambda^{-1}$ emissivity law), which obviously means that 
the 12 micron flux is not dominated by the cool interstellar dust 
(in agreement with the IRAS data from other S0 galaxies; see Soifer et al 
\cite{Soifer87}, and references therein). 
Using the relation between dust mass, infrared luminosity, and $T_D$ given by 
Soifer et al (their eq.\,1), we derive flux densities which are below the IRAS 
limit for $T_D < 40$\,K. Hence, the IRAS upper limits for $\lambda > 12\,\mu$m 
do not contradict to the estimated dust mass as an upper limit. A substantial 
fraction of the dust in a warm phase at $ T \approx 200$\,K, as was found e.g. 
in the centre of the S0 galaxy NGC\,3998 (Knapp et al \cite{Knapp96}), can be 
excluded for NGC\,3384.   

We emphasize that the above estimate of dust reddening is relatively 
uncertain. In particular, a correction is obviously required with regard 
to assumption (5): the dust obscures only the light 
sources that are behind it and not the stars between the dust and the observer. 
Moreover, the estimated values are further reduced by a factor of $\sim 3$ if 
we identify the size of the CO-emitting region in NGC\,3384 with the beam size 
of the observations ($21\arcsec$ FWHM) rather than the median value of the 
galaxy sample. 

A more fundamental question arises from the facts that dust 
(1.) is expected to be present as a consequence of the evolution of the stellar 
population and (2.) should be related to the amount of cool gas, namely:
why is NGC\,3384 that dust-poor?  
An explanation for the under-abundance of dust in S0 galaxies was proposed by
Lauer et al (\cite{Lauer05}). According to their scenario, dust 
appears at times throughout the galaxy and then falls to the centre where it is being 
processed or destroyed e.g., due to sputtering by hot X-ray gas. This assumption 
of episodic dust settling implies that dust is appearing and disappearing continually in 
early-type galaxies i.e., dust-free galaxies represent the end of a natural cycle where 
galaxies with lots of dust represent the other end. The length of the dust-free phase was 
estimated by Lauer et al to be on the order of a few times $10^7$\,yr and the length of 
the whole cycle is only $10^8$\,yr, i.e. much shorter than the timescale for the infall 
of gas from the outer parts of the galaxy or from outside.  Though this 
concept of a dust settling sequence appears attractive, a clear idea of the main driver 
of the cycle is still lacking, as was pointed out already by Lauer et al. 

Assuming the majority of the dust in the centre of NGC 3384 has been 
processed/destroyed, the question remains why molecular gas is detected there. 
If (1.) the dust cycle represents a cycle of accretion of matter by the central 
black hole and (2.) nuclear activity is signified 
by optical line emission -- as assumed by Lauer et al -- NGC\,3384 must be in 
a post-activity phase. Assuming that the cool ISM has been completely destroyed 
in the activity phase, by either the accretion onto or the feedback from the 
nuclear black hole, 
it can consist now only of the material returned by evolved stars 
during the dust-free phase of the cycle. Since the timescale of this phase is about 
$10^{-3}\,t_{\rm H}$, the mass of the cool ISM is expected 
to be $\la 10^{-3}$ times the mass returned during $t_{\rm H}$. 
This is indeed consistent with  
what was found for the most gas-poor S0 galaxies by Sage \& Welch (\cite{Sage06}). 
To check this idea further, we searched for a correlation bet\-ween the 
relative (i.e., observed {\it versus} predicted)
gas mass and the presence of dust in the centres of S0 galaxies. Unfortunately, 
there is no much overlap between the Welch \& Sage sample and the Lauer et al 
sample. Among the six galaxies which are in both samples, five have measurements 
of the gas mass, two of them show dust obscuration (NGC\,3607 and NGC\,4026). 
For the ratio of the observed gas mass to the estimated gas mass returned by 
evolved stars over $t_{\rm H}$, the mean value  for the dust-free galaxies is 
$10^{-3}$, compared to $0.035$ for the dusty galaxies. This seems to indicate, 
that dust-poor S0 galaxies indeed represent the phase of a generally reduced 
cool ISM. The statistics is however too poor to draw stringent conclusions. 
In the present context, we conclude that the estimated small dust reddening from 
the gas mass is at least not at variance with the scenario of the dust 
settling cycle. 
     
To summarize this discussion, we conclude that NGC 3384 is likely in the 
dust-poor phase of the dust-settling cycle. Nevertheless, dust reddening by  
$E(B-V) \approx 0.1$\,mag is not implausible in the central region of 
NGC 3384, though there is no direct clear-cut observational evidence.

%
\section{Conclusions}
%

Based on {\it HST} archive images we were able to confirm 
the compact inner component (IC) in the centre of NGC\,3384. 
This peculiar structure, discovered by Busarello et al 
(\cite{Busarello96}) on a Laplace-filtered NTT image, was hitherto
not seen on direct images. We visualize the IC using various methods 
of image processing. The images show also the extension of the well-known
bar-like elongated component (EC) toward the centre of the galaxy.  
For the IC, our results confirm the previously estimated maximum semi-major 
axis of $5\ldots6$\arcsec ($\sim 300$\,pc) and the position angle 
$PA \approx 46\degr$. The superb resolution of the {\it HST} images 
allows a better determination of the shape and the relative light contribution. 
The IC is found to contribute up to $\sim$40\% i.e., a factor of $\sim$8 higher than
estimated by Busarello et al. For the EC, on the other hand, we estimate
a slightly lower fraction than Busarello et al. This is in agreement with
the visual impression of the well-known peanut-shape because the 
maximum relative contribution of the EC is at larger distances, out of the
{\it HST} field. 
The maximum ellipticity of $\epsilon_{\mathrm{\, max, IC}} \approx 0.5$ 
is higher than determined in previous studies and signifies the same
inclination as the outer disk. The disk-like structure of the
IC seems to support the interpretation as an inner disk which is
in line with the kinematical data from two-dimensional spectroscopy
(Busarello et al \cite{Busarello96}; Fisher \cite{Fisher97}; Sil'chenko 
et al \cite{Silchenko03}; Emsellem et al \cite{Emsellem04}). The intensity
profile of the IC is not well approximated by a usual exponential law
but becomes shallower toward the centre.

Our broad-band colour indices from Calar Alto images
are shown to be in agreement with model predictions for a 5 to 7\,Gyr old 
stellar population of quasi-solar metallicity. No significant large-scale
variations of the colour indices over the main body of the galaxy are found.
Based on the population synthesis models from Salasnich et al 
(\cite{Salasnich00}), we discuss the expected colour differences 
between the EC/IC and the host galaxy. We find that it is not 
possible, based on our colour measurements, to discriminate between 
scenarios where galaxy components were made of the same young stellar 
population as the nucleus or by the older population of the host galaxy.  
The previously reported colour gradients close to the nucleus 
are most plausibly explained by small amounts of dust strongly 
concentrated to the centre, although NGC\,3384 appears to have, all in
all, a very low dust fraction. According to the episodic dust settling scenario 
suggested by Lauer et al (\cite{Lauer05}),
the low dust fraction is supposed to be an indication 
that NGC\,3384 is in a post-activity phase 
and at the beginning of a new dust-settling cycle.

\vspace{1cm}

\acknowledgements
H. Ismail is very grateful to TLS for kind hospitality and 
to the Deutsche Forschungsgemeinschaft for financial support.
Part of this work has been done using the {\it HST} data from the
data archive at the Space Telescope Science Institute and data
from the {\it Chandra} Data Archive. STScI 
is operated by the Association of Universities for Research in 
Astronomy, Inc., under NASA contract NAS 5-26555. 
The {\it Chandra} Data Archive is part of the {\it Chandra}
X-Ray Observatory Science Center which is operated for NASA
by the Smithsonian Astrophysical Observatory.


\end{document}